\documentclass[pre,floats,superscriptaddress,usenames,showpacs]{revtex4}
\usepackage{amsmath}
\usepackage{graphicx}
\usepackage[dvipsnames]{color}

\newcommand{\be}{\begin{equation}} 
\newcommand{\ee}{\end{equation}}
\newcommand{\bea}{\begin{eqnarray}}   
\newcommand{\eea}{\end{eqnarray}}

\newcommand{\rr}{{\bf r}}
\newcommand{\NN}{{\bf \nabla}}
\newcommand{\FF}{{\bf F}}
\newcommand{\GG}{{\bf G}}

\newcommand{\vv}{{\bf v}}
\newcommand{\vva}{{\bf v}^{\alpha}}
\newcommand{\vvb}{{\bf v}^{\beta}}
\newcommand{\vvab}{{\bf v}_{\alpha\beta}}
\newcommand{\fa}{f^{\alpha}}
\newcommand{\na}{n^{\alpha}}
\newcommand{\nb}{n^{\beta}}
\newcommand{\rhoa}{\rho^{\alpha}}

\newcommand{\fb}{f^{\beta}}
\newcommand{\cc}{{\bf c}}
\newcommand{\uu}{{\bf u}}
\newcommand{\uua}{{\bf u}^{\alpha}}

\newcommand{\uai}{u^{\alpha} _i}
\newcommand{\uaj}{u^{\alpha }_j}

\newcommand{\uub}{{\bf u}^{\beta}}
\newcommand{\ma} {m^{\alpha}}

\newcommand{\mb} {m^{\beta}}

\newcommand{\sab}{\sigma_{\alpha\beta}}
\newcommand{\gab}{g_{\alpha\beta}}
\newcommand{\muab}{\mu_{\alpha\beta}}

\newcommand{\bk}{{\bf k}}

\begin{document}
\date{\today}
\title{Dynamics of Fluid Mixtures in Nanospaces}

\author{Umberto Marini Bettolo Marconi\footnote[3]
{(umberto.marinibettolo@unicam.it)}
}

\address{ Scuola di Scienze e Tecnologie, 
Universit\`a di Camerino, Via Madonna delle Carceri, 62032 ,
Camerino, INFN Perugia, Italy}

\author{Simone Melchionna}

\address{Institute of Materials, Ecole Polytechnique F\'ed\'erale de 
Lausanne (EPFL), 
1015 Lausanne, Switzerland and Istituto Processi Chimico-Fisici, Consiglio Nazionale delle Ricerche, Italy}
\begin{abstract}

A multicomponent extension of our recent theory
of simple fluids [ U.M.B. Marconi and S. Melchionna, Journal of Chemical Physics, 131, 014105 (2009) ] is proposed to
describe miscible and immiscible liquid mixtures
under inhomogeneous, non steady conditions typical of confined
fluid flows. We first derive from a microscopic level
the evolution equations of the phase space distribution function of
each component in terms of a set of self consistent fields,
representing both body forces and viscous forces
(forces dependent on
the density distributions in the fluid and on the velocity
distributions).  Secondly, we solve numerically
the resulting governing equations  by means of the
Lattice Boltzmann method whose implementation contains novel
features with respect to existing approaches.  Our model incorporates
hydrodynamic flow, diffusion, surface tension, and the possibility for
global and local viscosity variations.  We validate our model by
studying the bulk viscosity dependence of the mixture on concentration,
packing fraction and size ratio. Finally we consider inhomogeneous systems and study the
dynamics of mixtures in slits of molecular thickness and relate structural and flow properties.

 \end{abstract}
\pacs{47.11.-j, 47.61.-k, 61.20.-p}

\maketitle

\section{Introduction}
In recent years, the experimental study of the flow of liquids in
nanometric structures has become an important and rapidly evolving
discipline driven by technological applications, such as lab-on-a-chip
systems, electrophoresis and electro-osmotic pumping.  The
mixing and separation of different molecules according to their size
or chemical and transport properties is a major issue in
chromatography, nanofluidic logic gates and electrokinetics.
Many applications are foreseen also in biotechnologies, because very
small quantities of liquid components are sufficient for analysis and synthesis,
including the possibility of isolating and analyzing or manipulating
a single molecule in a single nanochannel or surface with nanoscale
features \cite{sparreboom,schoch}.

These technological advances require a better theoretical
understanding of the fluid properties in confined geometries and under
non-equilibrium conditions.  At the nanoscale, new areas of physics,
chemistry and materials science come in.  The systems become more
surface-like and molecules never explore a bulk-like environment and because 
inhomogeneities have a great impact some interesting
phenomena arise.  Whereas a large body of information has been
accumulated in the last thirty years concerning the physics
of inhomogeneous fluids under thermodynamic equilibrium conditions
\cite{Evans2}, the state of the art of flowing fluids is not so
advanced \cite{nicholson,aiche,majumder,rauscher}.

It is crucial to understand how the interplay between the various
forces occurs and how it changes when going from macrosystems to
nanosystems.  It is clear from dimensional considerations that the
relative importance of forces changes with the typical dimensions of the systems
and we expect that the smaller the sample under scrutiny the more
important the role of surface forces is.  Not
only nanoscale flows are dominated by viscous effects if turbulence
is absent, but wherever frequent
molecule-channel wall collisions occur besides molecule-molecule
collisions, we expect a modified frictional resistance \cite{Guo1}.

%%%%%%%%%%%%%%%%%%%%%%%%%%%%%%%%%%%%%
%\section{Introduction}

The generalization of fluid dynamics from pure to multicomponent
fluid mixtures composed of different components or species
requires the introduction of some new concepts.  While traditional
computational fluid methods involve some {\it ad hoc} extrapolation of
Navier-Stokes equation to confining geometries, our approach is based
on a microscopic kinetic equation incorporating the effects of
the inhomogeneities in its structure.

The major difference with respect to one-component fluids is the
phenomenon of concentration diffusion.  We shall study a
multicomponent kinetic equation
\cite{Chapman,Degroot,Pina,Ferziger,Tham} that has been considered in
the past by several authors with the specific goal of computing the
transport coefficients from a microscopic approach.  Many of these
studies, which were based on the Boltzmann equation or on its
extension to the dense case, the Boltzmann-Enskog equation, require
some analytical and numerical effort, so that a simpler treatment of
the interactions has been proposed.  A very popular approximation is
represented by the phenomenological Bhatnagar-Gross-Krook (BGK)
equation whose merit is to reduce the complexity of the original
Boltzmann collision kernel $\Omega^{\alpha\beta}$ between species
$\alpha$ and $\beta$ by introducing a simple relaxation-time ansatz
\cite{BGK,Gross,Garzo,Andries,Sofonea}.  The BGK, in spite of being even less accurate
than the original Boltzmann equation, has enjoyed a great popularity
especially in conjunction with  the Lattice Boltzmann
method (LBM) thanks to the simplicity  of the collision kernel
leading to a considerable speed-up of the numerics \cite{LBgeneral} .

However, in the multicomponent case the choice of
the form of the BGK relaxation term is not unique and hard to infer
from the original $\Omega^{\alpha\beta}$, 
and  the literature abounds of several
proposals \cite{Hamel,Luo,Zhaoli,Asinari,Xu}.

 In order to describe with sufficient accuracy the fluid structure
at length scales comparable with the size of the particles we shall resort
to methods similar to those of density functional theory (DFT)
employed in the study of equilibrium and non equilibrium properties
\cite{Evans1,Tarazona,ArcherEvans,Lowen}.  In the case of hard-core
fluids,  DFT and its dynamical extension gives excellent results and
can be extended to more realistic fluids by using the van der Waals
picture of decomposing the total inter-particle potential into a
short-range repulsive potential and a long-range attractive potential
tail. The first is treated by means of a reference hard-sphere system
whilst the second is considered within the random phase 
approximation (RPA).  Such a decomposition can describe phase separating
systems, wetting phenomena, two phase interfaces, but predicts
transport coefficients of purely hard-core systems \cite{Espanol}.
 
The hard-core part of the interaction is treated within the revised
Enskog theory (RET) \cite{vanbeijeren} that considers the non-local
character of the momentum and energy transfer and takes into account
the static spatial correlations among  particles. This level of theory does not
incorporate the time correlations which are responsible for memory
effects and hydrodynamic contributions to the transport
coefficients \cite{Sung}.

With respect to the dynamical DFT \cite{umbmnogalilei,Cecconi}, the present
theory preserves  Galilei invariance   and therefore
displays full hydrodynamic behavior in the limit of slowly varying
fluctuations about the equilibrium reference state.

With the aim of deriving a practical numerical approximation to study
mixtures in inhomogeneous situations, we extended the method of Dufty
and coworkers \cite{Brey} to the multicomponent case.  The method is a
compromise between the RET, of which it retains the accuracy as far as
the momentum and energy transfer are involved, and the much simpler
BGK, which we employ to evolve the non-hydrodynamic moments of
the distribution functions \cite{Melchionna2008,Melchionna2009, Lausanne2009}.

The final product of our theory is a coupled system of simplified
equations for the density distributions of individual species
describing both the streaming and the collisional stages. These equations are 
solved by an appropriate extension of the LBM algorithm to include
both particle-particle interactions and particle-wall interactions.  A
simple analysis of the equations is used to derive explicit
expressions both for equilibrium thermodynamic quantities, such as
pressure, compressibility, etc., and for non equilibrium transport
coefficients.  It is important to stress the difference between the
present work and other LBM based approaches for non-ideal fluids,
where only attractive interactions are accounted for by the so-called
Shan-Chen pseudo-potential, whose justification is purely mesoscopic,
while transport coefficients enter as free parameters of the theory
\cite{shanchen,Shan}.

Interestingly, over two decades ago Davis and coworkers proposed on
phenomenological grounds a local average density model (LADM) of
viscosity and diffusivity, based on the principle of local averaging
of the density over the volume of a hard sphere \cite{LADM}.  The
present approach derives similar expressions for the transport
coefficients, but differs from the LADM in the way it solves the
resulting evolution equations.  Pozhar gave a more
rigorous foundation to the transport theory in non-uniform systems, but
the resulting theory could not be developed into a computationally
tractable method \cite{Pozhar, Pozhar2}.  In the present work, we claim that our
microscopically founded theory lends itself to relatively simple
numerical solutions based on the use of the LBM even in the presence
of complex geometries.
The numerical application regards fluids confined to a spacing of a few monolayers between 
solid surfaces, where they may behave very differently from 
bulk conditions and  exhibit unusual properties such as a dramatic increase in viscosity.

The paper is organized as follows: in Sec.
\ref{kineticequations} we introduce the model and the evolution
equations for the individual distribution functions of the
multicomponent system. and in Sec. \ref{Simplified}
we discuss the approximations involved in the theory.
In Sec. \ref{Channel} we test the predictions of the resulting  evolution equations 
using a  simple confining geometry.
 In Sec. \ref{Validation} we validate the
resulting equations numerically and simulate the flow of a mixture in narrow
channels.  Finally in Sec. \ref{Conclusions} we present our conclusions
and perspectives. Two appendices are added to the paper in order to 
illustrate the relationship between the collision kernel and the non-ideal part of the chemical potential
and discuss some technical details related to the  discretization procedure. 

%%%%%%%%%%%%%%%%%%%%%%
%%%%%%%%%%%%%%%%%%%%%

\section{Kinetic equations for the distribution functions}

\label{kineticequations}
We consider a fluid mixture of $N^\alpha$ particles of
mass $\ma$ , with $\alpha=1,M$, where $M$ represents the number of
components, mutually interacting with the potentials
$U^{\alpha\beta}(\rr,\rr')$. The probability density of finding a
particle of type $\alpha$ at position $\rr$ with velocity $\vv$ at
time $t$ is proportional to the the single-particle distribution
functions, $\fa(\rr,\vv,t)$, normalized as follows:
\be
N^\alpha=\int d\rr \int d\vv \fa(\rr,\vv,t) .
\label{normalization}
\ee
The evolution of $\fa$ is given, in principle, by the exact BBGKY
hierarchy of dynamical equations for the reduced distribution functions \cite{Chapman}, 
whose first member is the following kinetic equation for the $\alpha$ component:
\bea      
\partial_{t} \fa(\rr,\vv,t) +\vv\cdot\NN \fa(\rr,\vv,t)
+\frac{\FF^{\alpha}(\rr)}{\ma}\cdot
\frac{\partial}{\partial \vv} \fa(\rr,\vv,t)
= \sum_\beta\Omega^{\alpha\beta}(\rr,\vv,t) ,
\nonumber\\
\label{uno}
\eea
where 
$\FF^{\alpha}(\rr)$ is an external velocity independent force field acting on component $\alpha$ and
$\Omega^{\alpha\beta}$
represents the effect of the
interactions among the fluid particles of  type $\alpha$ and $\beta$.
In principle, the exact form of $\Omega^{\alpha\beta}$  is known, 
but it involves the two-particle distribution functions, which in turn depend on the three particle
distribution functions.  In order to close this hierarchy different approximations have been
proposed, the first historically being the Boltzmann equation expressing  $\Omega^{\alpha\beta}$  
in terms  of products of single-particle distributions.
Other specific choices of $\Omega^{\alpha\beta}$ in eq.(\ref{uno}) correspond to 
different kinds of approximations and lead to equations   such as the BGK
equation \cite{BGK}, the Vlasov  or RPA equation \cite{hansen} and the RET equation  \cite{vanbeijeren}.

 \subsection{ Macroscopic variables.}
 We seek a description in terms of a small number of macroscopic variables
which are functions of the fixed position $\rr$ and the time $t$.
The procedure leading from eq.(\ref{uno}) 
to the evolution equations in terms of  these variables,  closely follows the analysis of the one-component case 
\cite{Melchionna2009}
 %%%%%%
and allows to derive a set of hydrodynamic 
equations, in the case of a M-component fluid mixture.
In $D$ spatial dimensions, there exist $L=(D+M+1)$ conservation laws, for the mass of each species, 
 global momentum and energy. 
Thus, one can consider the evolution of   $L$ linear combinations of 
velocity moments of $\fa$ which are identified with the  hydrodynamic
fields of the mixture.

In the following, we shall adopt the Einstein summation convention for repeated Cartesian indexes,
while sums over components are written explicitly. 
Let us first write the statistical expressions for 
the mass density of  component $\alpha$:
\be
\rhoa(\rr,t)=\ma \na(\rr,t) = \ma \int d\vv \fa(\rr,\vv,t),
\label{density}
\ee
where $\na$ is the associated number density.
The total average mass density is given by:
\be
\rho(\rr,t)=\sum_{\alpha}\rhoa(\rr,t) ,
\ee
while the total number density  is $n\equiv\sum_\alpha \na$.
The first velocity moment of the phase-space distribution
corresponds to
the average local velocity, $\uua$,  of the $\alpha$ component
\be
\uua(\rr,t)=\frac{1}{\na(\rr,t)} \int d\vv \vv \fa(\rr,\vv,t).
\label{momentum}
\ee
We also consider the local barycentric velocity:
\be
\uu(\rr,t)=\frac{\sum_{\alpha}\rhoa(\rr,t)\uua(\rr,t)}
{\sum_{\alpha}\rhoa(\rr,t)}
\ee
and the so called dissipative diffusion current, 
measuring the drift  component $\alpha$ with respect to the center of mass velocity:
\be
{\bf J}^{\alpha}(\rr,t)
%=\ma \int d\vv \fa(\rr,\vv,t) (\vv-\uu)
=\rhoa(\rr,t)[ (\uua(\rr,t)-\uu(\rr,t)]
= \rhoa(\rr,t) {\bf w}^\alpha(\rr,t)
\ee
%where ${\bf w}^{\alpha}\equiv (\uu^\alpha-\uu)$ and 
and having the property $\sum_\alpha  {\bf J}^{\alpha}(\rr,t)=0$.

\subsection{Balance equations.}
The macroscopic variables $\rhoa,\uua,\uu$ obey a set of evolution equations derived from eq. (\ref{uno})
as follows.
Upon multiplying by $\ma$ and integrating over the velocity degrees of freedom both sides of
eq. (\ref{uno}) one obtains the continuity equation for the mass density of the species $\alpha$:
\be
\partial_{t}\rhoa(\rr,t) =-\nabla\cdot \Bigl( \rhoa(\rr,t) \uua(\rr,t)\Bigl)=-\nabla\cdot \Bigl(\rhoa(\rr,t) \uu(\rr,t)+ {\bf J}^{\alpha}(\rr,t) \Bigl) .
\label{continuity}
\ee
After summing over components, one obtains the global mass continuity equation:
\be
\partial_{t}\rho(\rr,t)+\nabla\cdot \Bigl( \rho(\rr,t) \uu(\rr,t)\Bigl)=0.
\label{densityconservation}
\ee 
In establishing eq. (\ref{continuity}) we have used the property that the
interaction does not affect the partial number densities, i.e.
$\int d\vv \Omega^{\alpha\beta}=0$.
The momenta of the single species are not  locally
conserved quantities, due to the interaction between different components. To derive the associated governing equations
we first multiply eq. (\ref{uno}) by $\ma (\vv-\uu)$ and then integrate over the velocity with the result:
\bea
&&
\partial_{t}[\rhoa(\rr,t)\uaj(\rr,t)]+ 
\nabla_i \Bigl(\rhoa(\rr,t) \uai(\rr,t) \uaj(\rr,t)
- \rhoa(\rr,t) w^{\alpha}_i(\rr,t) w^{\alpha}_j(\rr,t)\Bigl)=
\nonumber\\
&& 
-\nabla_i \pi_{ij}^{\alpha}+ \frac{F^{\alpha}_j(\rr)}
{\ma}\rhoa(\rr,t)+
\sum_{\beta} C^{\alpha\beta}_j(\rr,t) .
\label{momentpartial}
\eea
In analogy with pure fluids, in eq. (\ref{momentpartial}) we have introduced
the kinetic contribution to the partial stress tensor:
\be
\pi_{ij}^{\alpha}(\rr,t)=  \rhoa(\rr,t)<(v_i-u_i)(v_j-u_j)>
=\ma\int d\vv (v_i-u_i)(v_j-u_j)\fa(\rr,\vv,t).
\label{pressurekin}
\ee
Notice that  in order to ensure that $\pi_{ij}^{\alpha}$ is independent of a superimposed uniform
translation of the entire flowing fluid its definition contains the difference
$(v^\alpha_i-u_i)$.
In eq. (\ref{momentpartial}) we have also defined the projection of the collision term onto the velocity:
\be
{\bf C}^{\alpha\beta}(\rr,t)=\ma \int d\vv  \vv \Omega^{\alpha\beta}(\rr,\vv,t) ,
\label{cdefinition}
\ee
representing  the contribution to the stress tensor arising from interactions.
This integral will be explicitly computed in Sec. \ref{Interactions} for a specific model.
Finally,
the term $\rhoa w_i^\alpha w_j^\alpha$  in the l.h.s. of eq.  (\ref{momentpartial}),
has nothing to 
do with viscous effects, but represents
an additional stress of dynamical
origin due to the different average velocities of the two components 
\cite{Ramshaw} and vanishes in the pure fluid. 
Summing over species
eq. (\ref{momentpartial})  we obtain
the local conservation law for the total momentum,
$\rho \uu$, 
\bea
&&\partial_{t}[\rho(\rr,t)u_j(\rr,t)]+ \nabla_i
\left(\rho(\rr,t) u_i(\rr,t) u_j(\rr,t)]\right)\nonumber\\
&& =-\nabla_i \pi^{(K)}_{ij}+ \sum_{\alpha}\frac{F^{\alpha }_j(\rr)}{\ma}
\rhoa(\rr,t)+
\sum_{\alpha\beta} C^{\alpha\beta}_j(\rr,t)
\label{momentumcont}
\eea
with the total  kinetic component of the stress tensor given by:
\be
\pi^{(K)}_{ij}(\rr,t)= 
\sum_{\alpha}  \pi^{\alpha}_{ij}(\rr,t).
\label{pressurekin}
\ee
The set of eqs. (\ref{continuity}) and (\ref{momentumcont}) is not determined unless
additional relations specifying the pressure tensor
(\ref{pressurekin})
in terms of the selected macroscopic fields are given.
Hydrodynamics   at the Navier-Stokes level 
closes the equations by  assuming a simple linear relation between fluxes and
derivatives of the fields via phenomenological  transport coefficients  
(such as the viscosity and conductivity) whose values are not known a 
priori, but which must be taken from experiments. 
Kinetic theory, working at the level of the  non-equilibrium distribution functions, avoids 
such assumptions.
% and  provides a method of reduction from microscopic variables to macroscopic variables.
 %%%%%%%%%%%%  

\section{ Simplified kinetic model}
\label{Simplified}
In a nutshell,  our strategy consists in solving  the kinetic equations rather than the
balance equations (\ref{continuity}) and  (\ref{momentumcont}). It involves concepts borrowed from the lattice Boltzmann method (LBM) and from density functional theory (DFT).
The  LBM can be regarded as  an efficient approach to 
treat inhomogeneous systems and 
multiphase flows and provides an alternative to the solution of the Navier-Stokes equations.  Conditions at solid boundaries are 
relatively easily implemented using the LBM, making it well suited for 
handling complex geometries. 
The LBM becomes particularly useful to treat non-local effects, such as wetting layers, interfaces, layering etc.,
if  used in conjunction with a simplified kinetic approach suggested
by Dufty and coworkers in the case of one-component fluids \cite{Brey}.

To introduce the simplified kinetic model for mixtures, we refer to ref. \cite{Melchionna2009}, where we discussed
the simpler one-component case and reduced the
 full transport equation (\ref{uno}) to a simpler form.
 The extension to the multicomponent case requires some subtleties
which are needed in order to satisfy a 
key requirement of this approximation:  
the modified kinetic equation and the original one must share the same 
 equations for the partial densities and momenta, i.e. the distributions  must have the same low order projections.
Doing so, we are able to treat exactly the contribution of  the collision operator to the
hydrodynamic equations. As a result, we need  to  make  approximations only on 
that part  of the collision operator which acts on the higher order,  non-conserved modes. 
This latter contribution can be handled 
in the spirit of some cruder approximation, such as the  BGK-like approximation  \cite{Sofonea}.  Since the velocity distributions 
in the final equilibrium state are known it is easy to construct the BGK collision operator
in such a way that it verifies some basic properties 
of the original problem, namely particle and momentum conservation.
Our two-component extension  of the simplified kinetic equation reads:
\bea
&&\partial_{t}\fa(\rr,\vv,t) +\vv\cdot\NN \fa(\rr,\vv,t)
+\frac{\FF^{\alpha}(\rr)}{\ma}\cdot
\frac{\partial}{\partial \vv} \fa(\rr,\vv,t)\nonumber\\
&&= \frac{1}{k_B T} \frac{\psi^{\alpha}(\rr,\vv,t)}{\na(\rr,t)}  
 (\vv-\uu) \cdot  \sum_{\beta}  {\bf C}^{\alpha\beta}(\rr,t)
%\nonumber\\&&
-\omega[ \fa(\rr,\vv,t)- \psi^{\alpha}_{\perp}(\rr,\vv,t)] ,
\nonumber\\
\label{breytwo}
\eea
where  $\omega$ is a collision  frequency which depends on the two species
and
\be
\psi^{\alpha}(\rr,\vv,t) \equiv \na(\rr,t)[\frac{\ma}{2\pi k_B T}]^{3/2}\exp
\Bigl(-\frac{\ma(\vv-\uu(\rr,t))^2}{2 k_B T} \Bigl)
\ee
and
\bea
&&
\psi^{\alpha}_{\perp}(\rr,\vv,t) \equiv \psi^{\alpha}(\rr,\vv,t) \Bigl\{1+
\frac{\ma(\uua(\rr,t)-\uu(\rr,t))\cdot(\vv-\uu(\rr,t))}{k_B T}\Bigl\} .
\nonumber\\
\label{prefactor}
\eea

The proposed form  (\ref{breytwo}) differs from the existing literature about mixtures in a few aspects.
In the first place, the mixtures were treated either by including the full collision operator
or by assuming the BGK to be a valid replacement. 
The simplified kinetic method, to the best of our knowledge, was never applied to mixtures.
 Our  choice for the BGK term is based on a single
relaxation time for the two components and  represents perhaps the minimal model  
satisfying the conservation laws compatible with the Dufty and coworkers strategy.

A few comments are in order. 
First of all, eqs. (\ref{continuity}), (\ref{momentpartial})  and (\ref{momentumcont}) 
 are immediately recovered after multiplying
eq. (\ref{breytwo}) by $\ma,\ma (\vv-\uu)$ and
integrating over the velocity, $\vv$.
Second,
the coupling between the evolution equations of the two distributions $\fa$ occurs through
the nonlinear collision kernel and this fact is also reflected in the BGK representation of that kernel.
 Equation (\ref{uno})  contains two terms $\Omega^{\alpha\beta}$, one for each type of interaction, but 
in the modeling of the BGK relaxation time approximation it is convenient to choose a single collision term.
This assumption is valid not only because
the coupling between the two distributions is achieved via the barycentric velocity $\uu$ contained
in  $\psi^\alpha$, but also through the term ${\bf C}^{\alpha\beta}$.
%. Most 
%of the recent studies of mixtures have been based on the single BGK collision model \cite{Sofonea}.
%with $\psi^\alpha_{\perp}\to \psi^\alpha$ and ${\bf C}^{\alpha\beta}=0, B^{\alpha\beta}=0$.
Third,
the classical BGK approximation when applied to a one-component
fluid would involve the difference between the distribution $\fa$ and the
local  Maxwellian $\psi^\alpha$.
In the present case a non exponential  prefactor  in eq.  (\ref{prefactor}) has been introduced and  
serves as to "orthogonalize" the term $\omega [ \fa- \psi^{\alpha}_{\perp}]$  
to the
"true" collisional terms,  ${\bf C}^{\alpha\beta}$, so that the BGK does not lead 
to modifications of the hydrodynamic equations.
Also notice that  the first velocity moment of $\psi^\alpha$ is $\na \uu$, whereas the first moment
corresponding to $\psi^{\alpha}_{\perp}$ is $\na\uu^\alpha$.
The present ansatz reduces to our  one-component equation \cite{Melchionna2009}
in the limit $\psi^{\alpha}_{\perp}\to \psi^{\alpha}$. 
The above choice satisfies the indifferentiability principle which dictates that
when all physical properties of the species are identical, the total distribution
$f=f^A+f^B$ obeys the single species transport equation.

%%%%%%%%%%%%%%%%%%%%%%
\subsection{Interactions}
\label{Interactions}
We consider 
the interaction term featuring in eq. (\ref{uno}) 
which according to the BBGKY hierarchy can be written as
\be
\Omega^{\alpha\beta}(\rr,\vv,t)=
\frac{1}{\ma}\NN_v \cdot \int d\vv' \int dr'
\NN_r U^{\alpha\beta}    
(\rr-\rr')f_2^{\alpha\beta}(\rr,\rr',\vv,\vv',t)   ,
\label{coll}
\ee
where $f_2^{\alpha\beta}(\rr,\rr',\vv,\vv',t)$
represents the two-particle distribution function 
and the $U^{\alpha\beta}((\rr-\rr')$ are the pair potentials 
between particles of species $\alpha$ and $\beta$.

In general, the intermolecular potential, $U^{\alpha\beta}$, 
contains both a repulsive
and an attractive contribution. The  latter 
plays a major role in determining the thermodynamic properties of the
system.
% and is responsible for the multiphase behavior. 
%It is therefore 
%necessary to include attractive potential tails. 
In the following we shall separate the repulsive and the
attractive contributions 
by representing the short range repulsion between the particles 
through  repulsive hard sphere potentials, characterized by the three
diameters $\sigma_{AA}$, $\sigma_{BB}$ and $\sigma_{AB}=(\sigma_{AA}+\sigma_{BB})/2$
plus an attractive contribution:
\be
\Omega^{\alpha\beta}(\rr,\vv,t)=\Omega^{\alpha\beta}_{rep}(\rr,\vv,t)+
\Omega^{\alpha\beta}_{attr}(\rr,\vv,t) .
\label{opcoll}
\ee
The repulsive part is treated within the following
RET approximation \cite{Lopezdeharo}:
\bea
&&\Omega_{rep}^{\alpha\beta}(\rr,\vva,\vvb,t)
= \sab^2\int d\vvb\int 
d\bk\Theta(\bk\cdot \vvab) (\bk 
\cdot \vvab)\times\nonumber\\
&&\{ \gab
(\rr,\rr-\sab\bk,t) \fa (\rr,{\bar\vv}^\alpha,t)\fb (\rr-\sab\bk,{\bar\vv}^\beta,t) \nonumber\\
&& -
\gab(\rr,\rr+\sab\bk,t)\fa(\rr,\vva,t)\fb(\rr+\sab\bk,\vvb,t)\} ,
\label{collision}
\eea
where $\vv_{\alpha\beta}=(\vva-\vvb)$, while ${\bar\vv}^\alpha$ and ${\bar\vv}^\beta$ are scattered velocities given by
\bea
&&
{\bar\vv}^\alpha=\vva-\frac{2 \mb}{\ma+\mb}(\bk\cdot\vv_{\alpha\beta})\bk\nonumber\\
&&
{\bar\vv}^\beta=\vvb+
\frac{2 \ma}{\ma+\mb}(\bk\cdot\vv_{\alpha\beta})\bk .
\label{collrule}
\eea
The quantities $\gab(\rr,\rr\pm\sab\bk)$ are the three
hard sphere pair correlation functions evaluated when the particles
of species $\alpha$ and $\beta$  are at contact. 
The RET collisional term describes
the large  momentum binary exchanges due to close encounters
between the cores, whereas the attractive term  can be seen as a slowly varying 
average force, involving   small momentum exchanges. 
%%%%%%%%%%%%%%%%%%%%
In detail, the attractive tails of the potentials are treated in a
 mean field approximation, rewriting the second term of  the r.h.s. 
of eq.(\ref{opcoll})  by factorizing the two-particle distribution functions 
\be
f_2^{\alpha\beta}(\rr,\vv,\rr',\vv',t)
\approx\gab(\rr,\rr',t)\fa(\rr,\vv,t)\fb(\rr',\vv',t),
\label{factoriz}
\ee
that is, by neglecting velocity correlations completely:
\be
\Omega^{\alpha\beta}_{attr}(\rr,\vv,t)\approx
\frac{1}{\ma}\NN_v \cdot \int d\vv' \int dr'
\NN_r U_{attr}^{\alpha\beta}((\rr-\rr')\gab(\rr,\rr',t)\fa(\rr,\vv,t)\fb(\rr',\vv',t)   ,
\label{collattr}
\ee
where $\gab(\rr,\rr',t)$ is the inhomogeneous the pair correlation function of the reference hard-sphere system.
Integrating w.r.t. the velocity, we obtain
the RPA approximation for the attractive term:
\bea
&&
 \Omega^{\alpha\beta}_{attr}(\rr,\vv,t)=
-\frac{\GG^{\alpha\beta}(\rr,t)}{\ma}\cdot \NN_v \fa(\rr,\vv,t) ,
\label{omegaattr}
\eea
where we introduced the molecular fields
\be
\GG^{\alpha\beta}(\rr,t)=
- \int dr' \nb(\rr',t)\gab(\rr,\rr')\NN_r U_{attr}^{\alpha\beta}(\rr-\rr').
\ee

%%%%%%%%%%%%%%%%%%%%%%%%%AAAAAAAAAAAAAAAAAAAAAAAAAAA

In order to evaluate the interaction contribution to the momentum balance equation, we use the definition (\ref{cdefinition})  
together with eqs. (\ref{collision}) and (\ref{omegaattr}). After some algebra,
using explicitly the collision rules eqs.(\ref{collrule})
we obtain the following expression:
\bea
&&
C^{\alpha\beta}_i(\rr,t)=-2 \muab\sab^2 
\int d \vva  \int d\vvb \int d\bk \Theta(\bk\cdot \vvab)  (\bk \cdot \vvab )
(\bk\cdot\vv_{\alpha\beta} )\bk_i
\nonumber\\
&& 
g_{\alpha\beta}
(\rr,\rr+\sab\bk,t) \fa (\rr,\vva,t)\fb (\rr+\sab\bk,\vvb,t)  + G_i^{\alpha\beta}(\rr,t)\na(\rr,t)
\label{cinque}
\eea
where we have introduced the reduced mass 
$\muab=\frac{\ma \mb}{\ma+\mb}$.
To proceed further, we invoke a local equilibrium approximation
and replace  the distribution functions within the integrals  (\ref{cinque}) by local Maxwellians
\be
\fa(\rr,\vv,t)\simeq \phi^{\alpha}(\rr,\vv,t) 
\equiv n^{\alpha}(\rr,t)       
[\frac{\ma}{2\pi k_B T}]^{3/2}\exp
\Bigl(-\frac{\ma(\vv-\uua(\rr,t))^2}{2 k_B T} \Bigl)
\label{approdistrib}
\ee
and perform the integrals involved in eq. (\ref{collision}), by expanding $\uua(\rr,t)$ about  $\uu(r,t)$. We also expand
  the functions at point
 $(\rr+\sab)$  about their values at $\rr$ up to first order in the differences.
The procedure is completely analogous to the one followed in ref. 
\cite{Melchionna2009}
 and  leads to the final result:
\bea
&&{\bf C}^{\alpha\beta}(\rr,t)=-\sab^2 
\int d\bk \bk
g_{\alpha\beta}(\rr,\rr+\sab\bk,t)n_{\alpha}(\rr,t) 
n_{\beta}(\rr+\sab\bk,t)\times\nonumber\\
&&\Bigl( k_B T
- 2 \sqrt{\frac{2\muab k_B T}{\pi} } \bk\cdot
[\uub(\rr+\sab\bk,t)-\uua(\rr,t)]
\Bigl)
+\GG^{\alpha\beta}(\rr,t)\na(\rr,t) .
  \nonumber\\
%&&+\frac{\ma}{\ma+\mb} [T(\rr+\ss\bk,t)-T(\rr,t)] \Bigl)
\label{new11}
\eea
\subsection{Decomposition of the interaction term into specific force contributions. }

It is instructive to rewrite
the term featuring in formula (\ref{new11})  as a sum of more specific forces:
\be
\sum_\beta{\bf C}^{\alpha\beta}(\rr,t)=\na(\rr,t)\Bigl( \FF^{\alpha,mf}(\rr,t)+\FF^{\alpha,drag}(\rr,t)+\FF^{\alpha,viscous}(\rr,t) \Bigl) .
\label{splitforce}
\ee
We identify the force, $\FF^{\alpha,mf}$,  acting on the $\alpha$ particle at $\rr$ due to
the influence of all remaining particles, being the gradient of  the so-called potential of mean force: 
\be
\FF^{\alpha,mf}(\rr,t)=-k_B T\sum_\beta\sab^2 
\int d\bk \bk
g_{\alpha\beta}(\rr,\rr+\sab\bk,t)
n_{\beta}(\rr+\sab\bk,t)+\sum_\beta \GG^{\alpha\beta}(\rr,t)
\ee
a drag force with local character:
\be
F_i^{\alpha,drag}(\rr,t)=
-\sum_\beta \gamma^{\alpha\beta}_{ij}(\rr)[u_j^\alpha(\rr)-u_j^\beta(\rr)]
\ee
and a viscous force of non local character:
\be
F_i^{\alpha,viscous}(\rr,t)=
\sum_\beta  
\int d\rr'  H_{ij}^{\alpha\beta}(\rr,\rr') 
[u^\beta_j(\rr')-u^\beta_j(\rr)] .
\ee
We have further defined a friction tensor:
\be
\gamma^{\alpha\beta}_{ij}(\rr)\equiv\int d\rr' H_{ij}^{\alpha\beta}(\rr,\rr')
 \label{cccinqued}
\ee
and a viscosity kernel:
\bea
&&
H_{ij}^{\alpha\beta}(\rr,\rr')\equiv 2 \sqrt{\frac{2\muab k_B T}{\pi}} 
 g_{\alpha\beta}(\rr,\rr',t)
\nb(\rr',t)
%\nonumber\\&& 
\delta(|\rr'-\rr|-\sab) \frac{(r'-r)_i(r'-r)_j}{|\rr'-\rr|^2} ,
\label{visckernel}
\eea
whose relation with the transport coefficients will  become clear in the following.

The term $\FF^{\alpha,mf}$ is related to the gradient of the excess chemical potential
of species $\alpha$, $\mu_{int}^{\alpha}(\rr)$,
due to the interactions among the particles. 
In Appendix  A we show that in the case of slowly varying density profiles one has the result
\be
\FF^{\alpha,mf}(\rr,t)= -\NN \mu_{int}^{\alpha}(\rr,t).
\label{potchimico}
\ee

The drag term is
proportional to the velocity difference of the two species and is present  even in the absence
of velocity gradients. The viscous term, on the other hand, is proportional to 
velocity gradients.
%%%%%%%%%%%%%%%%%%%%%%%%%%%%%%%%%%%%%%%%% 
%{\it Frictional force}.
We can consider the case $\uu^A=const$ and $\uu^B=const$ and  $\uu^A\neq \uu^B$
and uniform densities so that the tensor $\gamma^{\alpha\beta}_{ij}$ becomes isotropic
\be
\FF^{A,drag}(\rr,t)=-\gamma^{AB} (\uu^A-\uu^B).
\ee
By using the definition (\ref{cccinqued}) we obtain
\be
\gamma^{AB}= \frac{8}{3}
(2 \pi \mu_{AB} k_B T)^{1/2}\sigma_{AB}^2 
g_{AB} n^{B} .
\ee
%%%%%%%%%%%%%%%%%%%%%%%%%%%%%%%%%
%This formula shows that the RPA term does not contribute to the
%transport coefficients, but only to the potential of mean force, whereas the hard core term does.

%%%%%%%%%%%%%%%%%%%%%%

%{\it Viscous force}.
It is interesting to show how the microscopic viscosity kernel (\ref{visckernel}) is related to the viscosity coefficient.
Let us assume a simple case $\uu^A=\uu^B=\uu$ with
\be
\uu(\rr,t)=(0,0,u_z(x,y))
\ee
and uniform densities, $n^A$ and $n^B$.
In this case the only non vanishing component of the viscous force is directed along $z$.
After some lengthy, but straightforward algebra (see ref \cite{Melchionna2009}) we obtain:
\be
F_z^{\alpha,viscous}
=\frac{4}{15}\sum_\beta\sqrt{2\pi\muab k_B T} \nb\sab^4\gab\Bigl(  \frac{\partial^2 u_z}{\partial x^2}+ \frac{\partial^2 u_z}{\partial y^2}  \Bigl)
\ee
which allows to identify the collisional contribution to the shear viscosity coefficient by using the macroscopic relation
\be
\sum_\alpha \na F_z^{\alpha,viscous}=\eta^{(C)}\Bigl(  \frac{\partial^2 u_z}{\partial x^2}+ \frac{\partial^2 u_z}{\partial y^2}  \Bigl)
\ee
from which we can write the explicit formula:
\be
\eta^{(C)}= 
\sum_{\alpha\beta}\eta^{(C)}_{\alpha\beta}
 =\frac{4}{15}\sum_{\alpha\beta}\sqrt{2\pi\muab k_B T} \na\nb\sab^4\gab.
\label{viscocollisional}
\ee
In the case of pure fluids, 
this expression of the viscosity  reduces to the one  predicted by Longuet-Higgins
\cite{Longuet}, which is known to describe successfully  fluids of hard molecules even when the radial distribution function is momentarily isotropic.
 In spite of the fact that the trial distributions correspond the the local equilibrium state,  
the singular nature of the interactions allows a finite flux of momentum.
Notice that the RPA attractive term does not contribute to the
transport coefficients, but only to the potential of mean force.

%\subsection{Nuovi integrali 9/2/2010}

%%%%%%%%%%%%%%%
%%%%%%%%%%%%%
\subsection{Fischer-Methfessel prescription for the pair correlation functions of a non uniform system.}
We have deduced the formulas for the viscosity and drag coefficient in the case of systems 
of uniform densities. In order to apply our theory to non-uniform cases we need to
specify how the pair correlation functions are modified with respect to the bulk.
The pair distribution function $\gab$ appearing in the above formulas is constructed according
to a generalization of the prescription put forward by 
Fischer and Methfessel (FM) \cite{Fischer}. One first defines the smeared densities
$\bar \na(\rr)$ via a uniform averaging of $\na(\rr)$ over spheres of volume
$V_{\alpha}=\pi \sigma_{\alpha\alpha}^3/6$
 centered at the point ${\bf R}_{\alpha\beta}$ located between the centers of the two
spheres at $\rr_\alpha$ and  $\rr_\beta$, respectively:
\be
{\bf R}_{\alpha\beta}=
\frac{\rr_\alpha+\rr_\beta }{2} .
\ee
The smeared densities are \cite{Sokolowski}:
\be
\bar \na({\bf R}_{\alpha\beta})=\frac{1}{V_\alpha}\int_{V_\alpha}d\rr 
\na(\rr-{\bf R}_{\alpha\beta}).
\ee
The
contact values of the pair correlation functions are computed using
the  bulk expressions
obtained from an extension \cite{Boublik,Mansoori}  of the  Carnahan-Starling equation to mixtures:
\be
\gab^{bulk}(\{ \xi_n\})
=\frac{1}{1-\xi_3}+\frac{3}{2}
\frac{\sigma_{\alpha\alpha}\sigma_{\beta\beta}}{\sab}\frac{\xi_2}{(1-\xi_3)^2}
+\frac{1}{2} \Bigl(\frac{\sigma_{\alpha\alpha}\sigma_{\beta\beta}}{\sab}\Bigl)^2
\frac{\xi_2^2}{(1-\xi_3)^3},
\label{carnahan}
\ee
where the functions $\xi_n$ are evaluated  for the values of the smeared densities
at point ${\bf R}_{\alpha\beta}$ between the two particles at contact:
\be
\bar \xi_n({\bf R}_{\alpha\beta}) =\frac{\pi}{6}\sum_{\alpha}\bar n^\alpha
({\bf R}_{\alpha\beta}) \sigma_{\alpha\alpha}^n.
\ee
The pair correlations at contact are evaluated in the inhomogeneous system using the generalized FM ansatz:
\be
\gab(\rr_\alpha,\rr_\beta;|\rr_{\alpha}-\rr_{\beta}|=\sab)=
\gab^{bulk}( \{\bar \xi_n({\bf R}_{\alpha\beta})\}).
\ee
From the knowledge of $\gab$, one can immediately obtain the equation of state for the mixture:
\be
\frac{P}{k_B T}=\na+\nb+\frac{2}{3}\pi\sum_{\alpha\beta}\na\nb \gab^{bulk}\sab^3.
\ee

After having made the connection  between the collisional term ${\bf C}^{\alpha\beta}(\rr,t)$
and the three types of forces acting in the fluid eq. (\ref{splitforce}), we can use the large body of knowledge accumulated over the last 
twenty years about entropic forces in multicomponent systems \cite{Likos}.
These forces, for instance,  determine an effective attraction towards a flat wall of the particles with the larger radius
induced by the presence of the small particles. 
For similar  entropic reasons two large particles immersed in a sea of small particles experience an effective attraction.
To the best of our knowledge, 
the interplay between entropic forces and viscous forces is not very well known and is worth to be explored.
As an order of magnitude estimate entropic forced are typically  $k_B T n\sigma_{AB}^2$, whereas
the ratio between viscous forces and entropic forces is $\approx \uu/v_T$ with $v_T=\sqrt{k_B T/\muab}$.

Although recent versions of equilibrium DFT are more accurate than 
the FM method, it is more convenient to work within the FM approximation for a series of reasons:
a) the FM approximation is very simple as compared to the Rosenfeld \cite{Rosenfeld}
prescription for pair correlations; b) our theory not only requires the effective potential $\nabla \mu_{int}^\alpha$,
but also the viscous and frictional forces which are not given by the DFT method; c) computationally the FM method is much faster, although it becomes 
numerically unstable at high packing fractions. 

\subsection{Numerical validation}
We consider a binary mixture of HS with $\sigma_{AA}=4$, $\sigma_{BB}=8$ and composition $X_A=n^A/(n^A + n^B)$.
The first validation of the numerical algorithm, succinctly described in Appendix B, is obtained by calculating the shear viscosity 
of the binary system and initially subjected to a sinusoidal shear wave of small amplitude 
in bulk conditions. 

\begin{figure}[h]
\begin{centering}
\includegraphics[width=12cm,clip=true,angle=0]{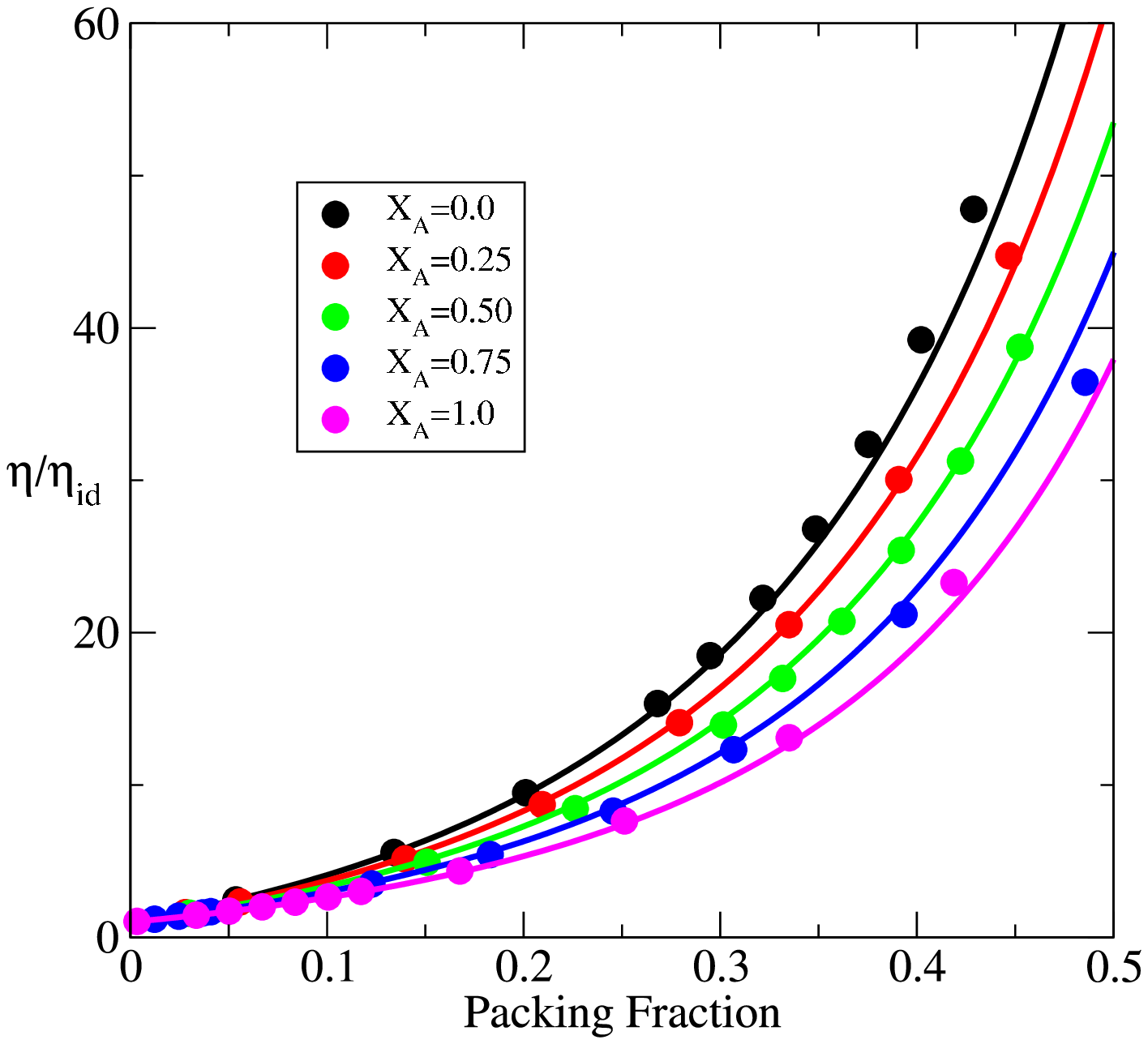}
\end{centering}
\caption{Shear viscosity (normalized to the ideal gas value) versus packing fraction for a bulk mixture at different values of the composition $X_A=n^A(n_A+n_B)$
of hard spheres of diameter ratio $\sigma_{BB}/ \sigma_{AA}=2$ .
The viscosity decreases with the concentration $X_A$. The continuous curves represent the values obtained with the theoretical formula, whereas the symbols
refer to the numerical values obtained from our LBM computer simulations.}
\label{fig:viscosity}
\end{figure}

The measured decay time is found to have the same temporal decay for both species, 
and the resulting viscosity curves are reported in Fig. \ref{fig:viscosity}. 
The data are compared with the analytical expression for the decay time obtained by solving the linearized hydrodynamic  equations
and using the theoretical viscosity of eq. (\ref{viscocollisional}). Notice that the viscosity increases monotonically with the packing fraction $\xi_3$ of
the system and depends on the molar fraction $X_A$. At fixed packing fraction, the larger the concentration of the big particles the larger is the viscosity.

The agreement between numerical and analytical data is remarkable. 
At large packing fraction, a degree of departure from the analytical data indicates the major role
played by the collisional integral that apparently is not fully captured by the
numerical quadratures. 
This is an effect of the discretization and of 
the linear interpolation employed in order to compute some off-lattice points used in the calculation of the surface
integrals.
% Overall, the systematically larger viscosity indicates that the 
% collisional term enhances interparticle collisions. 
From  the viewpoint of numerical stability, 
the range of packing fraction handled in simulation shows that the method is stable up to 
packing fractions of $\xi_3 < 0.4$, 
rendering the approach usable in many experimental condensed matter contexts. In
proximity of hard walls and under strong flow conditions, the numerical stability range reduces
to packing fraction below $\xi_3 \sim 0.3$.

%%%%%%%%%%%%%%%%%%%%%%%%%%%%%%%%%%%%%%%
\section{Flow in a slit-like pore. Toy model.}
\label{Channel}

\begin{figure}[htb]
\includegraphics[clip=true,width=12.0cm, keepaspectratio,angle=0]{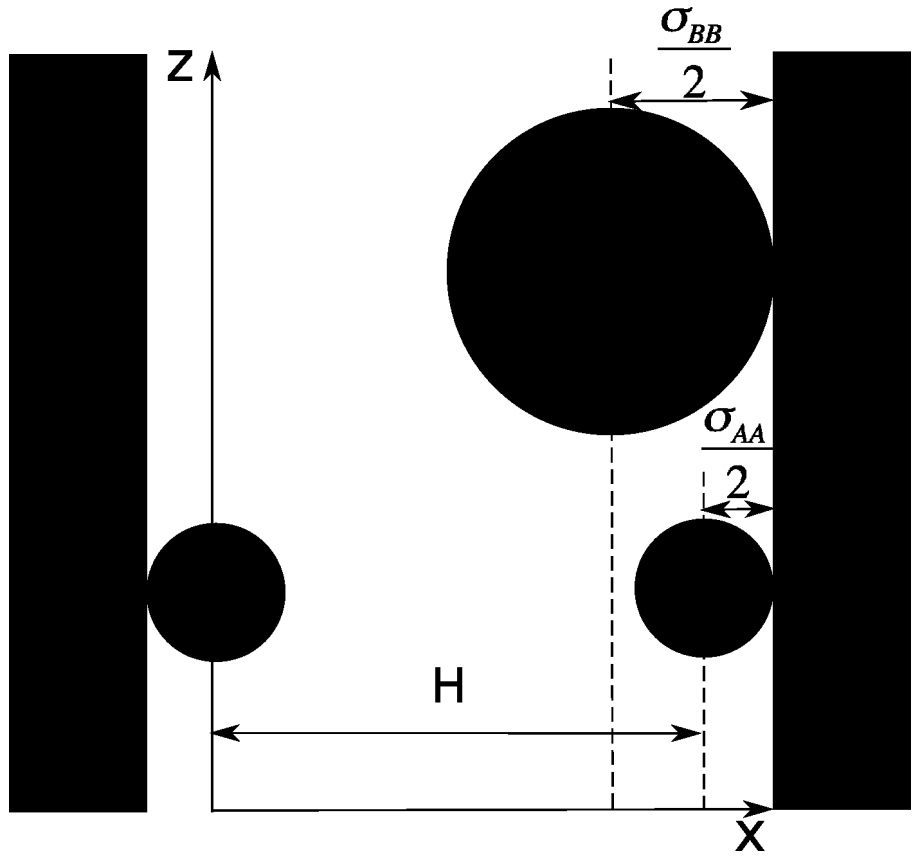}
\caption{ Sketchy view of the system.  The walls, parallel to the $yz$ plane are located at positions $x=-\sigma_{AA}/2$
and  $x=H+\sigma_{AA}/2$, the flow occurs along the vertical $z$ direction. The direction $y$ is normal to the
figure.}
\label{fig:geometry}
\end{figure}

We now  consider  a binary mixture of particles of equal masses and different sizes ($\sigma_{AA} <\sigma_{BB}$) confined in a  narrow slit-like pore
represented by two parallel smooth plates having infinite area and separated by a distance $H+\sigma_{AA}$
(see fig. \ref{fig:geometry}). The walls are parallel to the $yz$ plane and a flow along the $z$ direction is induced by the
presence of a uniform external field, $\FF$, also directed along $z$.
Since $H$ is the typical size of the system, $u$ the average flow velocity and $\eta$ the viscosity,
the Reynolds number is $ Re=\rho u H/\eta$. In the small $Re$ regime,   flows have negligible inertia
forces. The viscous and pressure forces should be in approximate balance and  a reduction, similar to the one
occurring between the macroscopic
Navier-Stokes and the Stokes equation, takes place also at the present microscopic level of description.

The  first remark is that, besides being small in systems  of microscopic size
and in the limit of low flow velocities,
the non linear term in the momentum equation vanishes. In fact, if one imposes the boundary conditions typical of a Poiseuillle flow, in 
a straight channel,
it vanishes due to symmetry.

 We rewrite the l.h.s. of eq. (\ref{momentpartial}) as:
\bea
&&
\partial_{t}[\rhoa(\rr,t)\uaj(\rr,t)]+ 
\nabla_i \Bigl(\rhoa(\rr,t) \uai(\rr,t) \uaj(\rr,t)
- \rhoa(\rr,t) w^{\alpha}_i(\rr,t) w^{\alpha}_j(\rr,t)\Bigl)
\nonumber\\
&&
=\uaj(\rr,t)\Bigl[\partial_{t}\rhoa(\rr,t)+\nabla_i(\rhoa(\rr,t) \uai(\rr,t) )
\Bigl]
+\rhoa\Bigl[\partial_{t}\uaj+\uai(\rr,t) \nabla_i\uaj(\rr,t)\Bigl]
-\nabla_i[\rhoa w_i w_j] .
\nonumber
\eea
The first parenthesis in the r.h.s. is zero due to the continuity equation of species $\alpha$, the second parenthesis is
also zero due to the geometry of the problem and the stationarity of 
the flux, and the last term vanishes because ${\bf w}$ is directed along $z$ 
and $\rhoa$ varies only along $x$.

The  flow induced by the presence of uniform fields parallel to the walls is characterized by velocities along the $z$ direction
$\uua=(0,0,u^\alpha_z(x))$ obeying the following set of stationary equations for each species:
\bea
&&
\sum_\beta  
\int d\rr'  H_{ij}^{\alpha\beta}(\rr,\rr') 
[\uub_j(\rr')-\uub_j(\rr)] -\frac{1}{\na}\nabla_j \pi^\alpha_{ij}=
\nabla_i \mu_{int}^{\alpha}(\rr)
+\sum_\beta \gamma^{\alpha\beta}_{ij}(\rr)[\uua_j(\rr)-\uub_j(\rr)]
-F_i^\alpha.
%\nonumber\\&&
\nonumber\\
\eea
In order to gain insight into the problem,
we first  set up and solve a toy model, obtained from the full model under some crude approximation.
We replace the true non-uniform density profiles by two slabs of constant densities, $n_A,n_B$, respectively:
\bea
n^A(x)&=&n_A\theta(x)\theta(H-x)\\
\nonumber
n^B(x)&=&n_B\theta(x-a)\theta(H-a-x) .\\
\eea
As shown in ref. \cite{Melchionna2009} the kinetic contribution to the viscosity due to the individual species is
$ \eta^{(K)}_{AA}=\frac{k_B T}{\omega} n_A$,
while the collisional part is given by formula (\ref{viscocollisional}) .
Notice that the above prescription for the kinetic contribution
to the viscosity is a consequence of our single time relaxation ansatz.
Let us introduce the abbreviations:
\be
\eta_{AA}=
\Bigl(\eta^{(K)}_{AA}+\eta^{(C)}_{AA}\Bigl) \, ,
\qquad
\eta_{BB}=
\Bigl(\eta^{(K)}_{BB}+\eta^{(C)}_{BB}\Bigl),
\qquad\qquad 
\tilde \eta_A= (\eta_{AA}+\eta^{(C)}_{BA} )
 \, ,
\qquad
\tilde \eta_B= (\eta_{BB}+\eta^{(C)}_{AB} ).
\ee

When the two species have different  diameters, 
the smaller species can approach the wall up to a distance
$\sigma_{AA}/2$, whereas the larger $B$ species can only reach the
distance $\sigma_{AA}/2+a$, where $a=(\sigma_{BB}-\sigma_{AA})/2$. In the following, we take $H>2a$ and
measure distances, $x$, from the position of closest approach of particles
$A$ to the wall as illustrated in Fig. \ref{fig:geometry}.

We now determine the velocity profiles consistent with our toy model.
In the central region $a<x<(H-a)$ we obtain the following coupled differential equations
for the partial velocities:
\bea
&&
\eta_{AA} \partial_x^2 u^A_z(x)
+\eta_{AB} \partial_x^2 u^B_z(x)=
\Gamma [u^A_z(x)-u^B_z(x)]
-F_z^A n^A \\
&&
\eta_{BB} \partial_x^2 u^B_z(x)
+\eta_{AB} \partial_x^2 u^A_z(x)=
-\Gamma [u^A_z(x)-u^B_z(x)]
-F_z^B n^B
\label{coupledvelo}
\eea
with 
\be
\Gamma= n^A \gamma^{AB}=\frac{8}{3}\sqrt{2 \pi K_B T\mu_{AB}} \sigma_{AB}^2  g_{AB} n^A n^B .
\ee
In the excluded layer for the big spheres, that is  for $x<a$ or $x>(H-a)$, the equations read: 
\be
\eta_{AA} \, \partial_x^2 u^A_z(x)= -F_z^A n^A
 \, ,
\qquad
u_z^B(x)=0.
\ee
%%%%%%%%%%%%%%%%%%%
%%%%%%%%%%%%%%%%%%
The solution 
with boundary conditions of zero velocity $u_z^A$ at $x=0$ and $x=H$, together with zero velocity $u_z^B$ at $x=a$ and $x=H-a$,
and continuous $u^A_z$ at $x=a$,
 reads:
\be
u^A(x)=\frac{F_z^A  n^A}{\eta_{AA}}\frac{(H-x)x}{2}, \quad   x<a, \quad x>(H-a)
\ee
while for $a<x<(H-a)$ is given by:
\bea
u_z^A(x)&=&\frac{F_z^A  n^A+F_z^B n^B}{\tilde \eta_A+\tilde \eta^B} \bigl(\frac{(H-x)x}{2}-
\frac{(H-a)a}{2}\bigl)+
C_A  \bigl(\cosh(\lambda(x-H/2))-\cosh(\lambda (a-H/2))\bigl)
\nonumber\\
&+&\frac{F_z^A  n^A}{\eta_{AA}}\frac{(H-a)a}{2}
\eea
and
\be
u_z^B(x)=\frac{F_z^A  n^A+F_z^B n^B}{\tilde \eta_A+\tilde \eta^B}\Bigl( \frac{(H-x)x}{2}-\frac{(H-a)a}{2} \Bigl) +
C_B \bigl(\cosh(\lambda(x-H/2))-\cosh(\lambda (a-H/2))\bigl)
\ee
with
\be
\lambda^2=\Gamma \Bigl(\frac{\tilde\eta_A +\tilde\eta_B}{\eta_{BB}\eta_{AA}
-(\eta_{AB})^2} \Bigl) 
\label{lambda}
\ee
and
\be
C_A=
\frac{\tilde \eta_B}{\tilde \eta_A+\tilde \eta_B}
\Bigl(\frac{1}{\Gamma}\frac{F_z^A n^A\tilde \eta_B-F_z^B n^B \tilde\eta_A}{\tilde \eta_A+\tilde \eta_B}
+\frac{F_z^A  n^A}{\eta_{AA}}\frac{(H-a)a}{2 } \Bigl)   \frac{1}{\cosh(\lambda(a -H/2))}
\ee
\be
C_B=
-\frac{\tilde \eta_A}{\tilde \eta_B} C_A.
\ee

\begin{figure}[htb]
\includegraphics[clip=true,width=12.0cm, keepaspectratio]{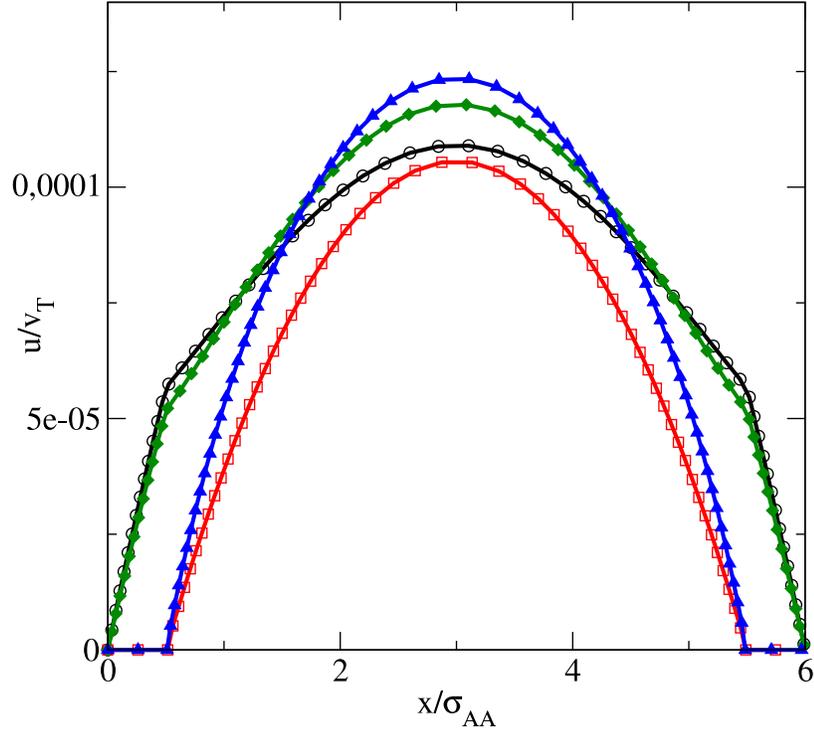}
\caption{Velocity profiles of the two species for a channel of width $H=6 \sigma_{AA}$ and load $F=0.001 k_B T/\sigma_{AA}$,
according to the toy model. 
The ratio of the diameters of the spheres   $\sigma_{BB}/\sigma_{AA}=2$.
Velocity of  species A  (open black circles)
and of species  B (red squares) at bulk composition $X_A=0.25$ and packing fraction $\xi_3=0.084$.
Velocity profile of species A  (green diamonds) and of species B (blue triangles ) at bulk composition $X_A=0.75$ and packing fraction $\xi_3=0.073$.}
\label{fig:doppioprofile}
\end{figure}

The main result of this illustrative model is the presence of different velocity profiles 
for the two species. The velocity profile of the bigger particles is generally lower than the corresponding profile of the smaller particles,
because B effectively sees a narrower slit, so that the confinement effect resulting in a Poiseuille parabolic-like profile
is more relevant than for the other species. The velocity difference between species remains finite  in the center of the pore.
Fig. \ref{fig:doppioprofile} displays the velocity profiles of the two components for two different values of the
composition and at low packing fraction. We will see below that even such a simplified model is able to capture some important
features which are better studied by means of our full numerical method.

%Finally, we notice that
%when $\lambda^2$ in eq. (\ref{lambda}) changes sign and becomes negative,  the behavior of the solution changes and 
%we observe the onset of an oscillatory behavior of the solution, that we interpret as the  lane formation instability.

\section{LBM numerical results}
\label{Validation}

In the present section we solve numerically the equations for the phase space distributions
and study the associated steady state properties. 
The most interesting applications of our method are those which regard truly inhomogeneous situations 
where  one observes the interplay between the microscopic structure and the 
flow properties.
The small amount of material in the confined region 
makes experimental probing of  confined 
fluid extremely difficult and thus numerical methods are welcome.

\subsection{ Profiles.} We reconsider the Poiseuille problem of the previous section using the full power of our theory to 
obtain the density and velocity profiles of the two components when a uniform field parallel to the
plates is applied.

The available region for each hard sphere can be computed according to the density profiles that
show where the centers of the spheres can range. Given that the sphere radii are multiples of the mesh spacing,
the extrema of the available region can only fall on a mesh point. In this respect, the values of $H$ are equal 
to $7, 13$ and $29$ mesh-points for the three channels, that is $H/\sigma_{AA}=1.75, 3.25$ and $7.25$ respectively.
On the other hand, for the velocity profiles, we have added a mesh spacing on the
left and right extrema to account for the non-slip boundary condition as imposed via
the bounce-back scheme in the numerical method. When looking at the intervals from
the velocity point of view, they look wider.

\begin{figure}[htb]
\includegraphics[clip=true,width=12.0cm, keepaspectratio,angle=0]{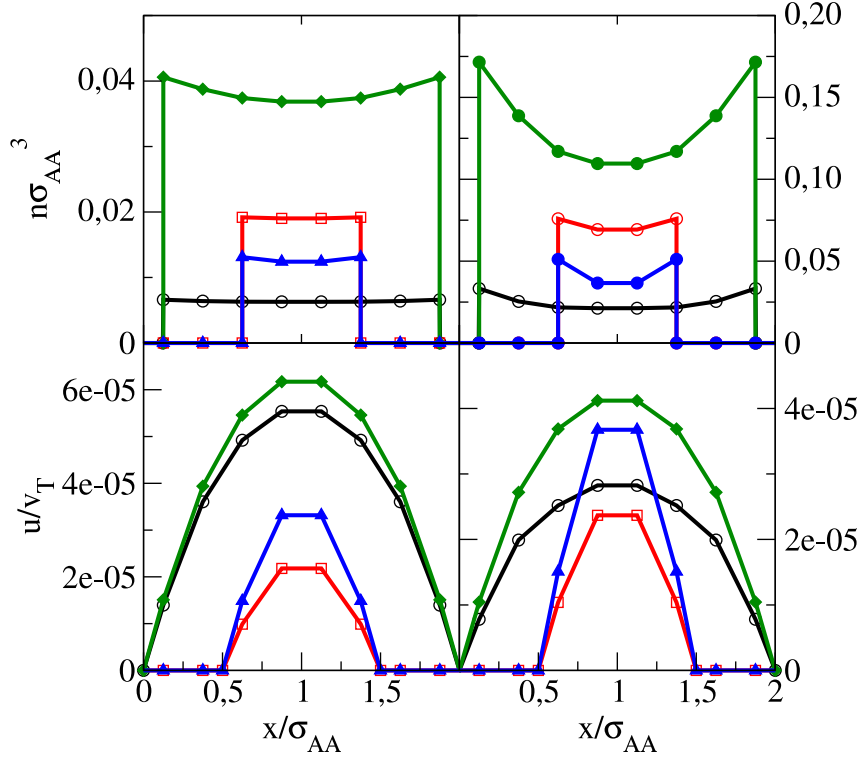}
\caption{Numerical results for a channel of width $H/\sigma_{AA}=1.75$ and load $F=10^{-6}k_B T/\sigma_{AA}$. The spheres ratio of
diameters is  $\sigma_{BB}/\sigma_{AA}=2$.
Upper left panel: Density profile of species A  (open black circles)
and of species B (red squares) at bulk composition $X_A=0.25$ and packing fraction $\xi_3=0.084$.
Density profile of species A  (green diamonds) and of species B (blue triangles ) at bulk composition $X_A=0.75$ and packing fraction $\xi_3=0.073$.
Lower  left panel: Velocity of  species A  (open black circles)
and of species  B (red squares) at bulk composition $X_A=0.25$ and packing fraction $\xi_3=0.084$.
Velocity profile of species A  (green diamonds) and of species B (blue triangles ) at bulk composition $X_A=0.75$ and packing fraction $\xi_3=0.073$.
Upper right panel: Density profile of species A  (open black circles) and
and of species B (red squares) at bulk composition $X_A=0.25$ and packing fraction $\xi_3=0.42$.
Density profile of species A  (green diamonds) and of species B (blue triangles ) at bulk composition $X_A=0.75$ and packing fraction $\xi_3=0.26$.
Lower  right panel: Velocity of  species A  (open black circles)
, of species  B (red squares) at bulk composition $X_A=0.25$ and packing fraction $\xi_3=0.42$,
velocity of species A  (green diamonds) and of species B (blue triangles ) at bulk composition $X_A=0.75$ and packing fraction $\xi_3=0.26$.}
\label{fig:narrow}
\end{figure}

Fig. \ref{fig:narrow} displays results for 
the density and  velocity profiles for each species in a channel of width $H=1.75 \sigma_{AA}$ 
for a mixture of particles with diameters $\sigma_{AA}=4$ and $\sigma_{BB}=8$ in two different cases: packing fraction $\xi_3=0.084$
and composition $X_A=0.25$ and     packing fraction $\xi_3=0.073$ and composition $x_A=0.75$.
In the upper left panel the mixture is almost a gas and the density profiles for both compositions considered do not display any significant structure,
whereas in the lower left panel the velocity profiles have nearly parabolic shapes, each species having its own curvature.
One may conclude that in such a regime the flow behavior of the confined fluid is Poiseuille-like. As we increase the packing fraction 
to $\xi_3=0.42$ (composition $X_A=0.25$)  and packing fraction $\xi_3=0.26$ (composition $X_A=0.75$), we observe
the behavior reported in the right panels of Fig. \ref{fig:narrow}. The large species $B$ is unable to display significant peaks, due to
the small plate separation that inhibits the formation of two layers of $B$ particles. The small species, instead, starts developing some 
density enhancement in the proximity of the walls (upper right panel). The velocity profiles instead  still bear a parabolic behavior with 
the smaller species always being  faster than the bigger species.

\begin{figure}[htb]
\includegraphics[clip=true,width=12.0cm, keepaspectratio,angle=0]{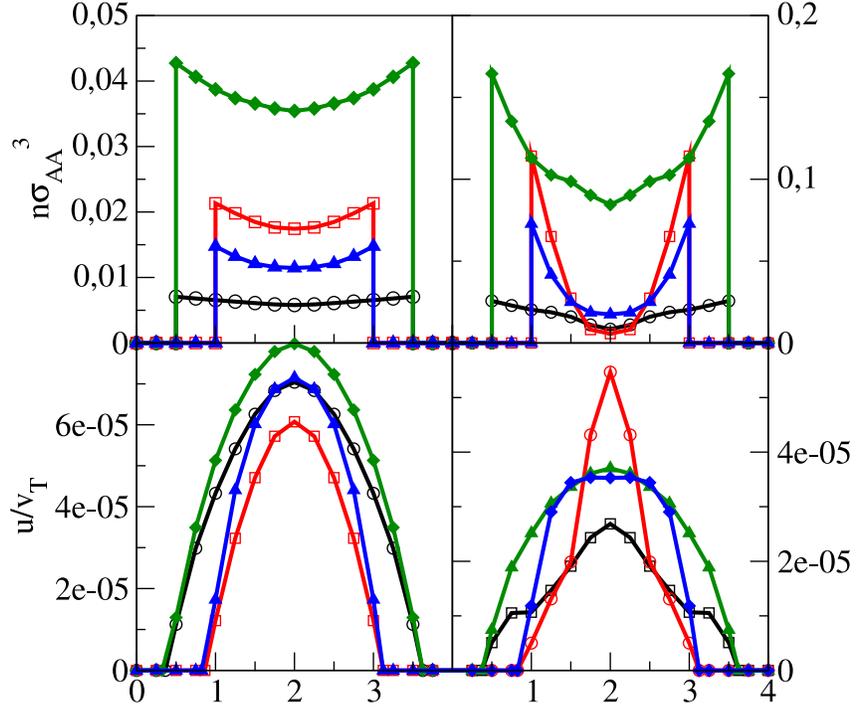}
\caption{Numerical results for a channel of width $H/\sigma_{AA}=3.25$ and load $F=10^{-6}k_B T/\sigma_{AA}$. The ratio of the 
diameters of the spheres is $\sigma_{BB}/\sigma_{AA}=2$ .
The symbols used in the four panels correspond to the same values of packing and concentration as those employed in figure \ref{fig:narrow}.}
\label{fig:medium}
\end{figure}

In Fig. \ref{fig:medium} we consider the same state conditions but a wider slit, $H=3.25\sigma_{AA}$, such that more than one
layer of particles B 
can fill the gap.
In the low packing region, at $\xi_3=0.073$ and at $\xi_3=0.084$  (upper left panel), both density profiles are rather
flat and the associated velocity profiles do not display significant structure, the
smaller particles being faster (lower left panel). 
As we  consider larger packing fractions, $\xi_3=0.42$  and composition $X_A=0.25$, we observe a stronger structure
in the large particles than in the small particles. The other
packing fraction, $\xi_3=0.26$ (composition $X_A=0.75$), also produces some significant peak structure at the walls
for the large particles, but the small particles have a rather flat density profile.
The small HS are somehow enslaved by the large component as far as the structure is concerned. 
In the velocity profiles (lower right panel) we observe that the $B$ component for $X_A=0.25$ has a higher velocity than the small 
counterpart near the center-line of the slit.
This effect is determined by  packing: small particles have more room near the walls while large particles move faster near the center-line.

\begin{figure}[htb]
\includegraphics[clip=true,width=12.0cm, keepaspectratio,angle=0]{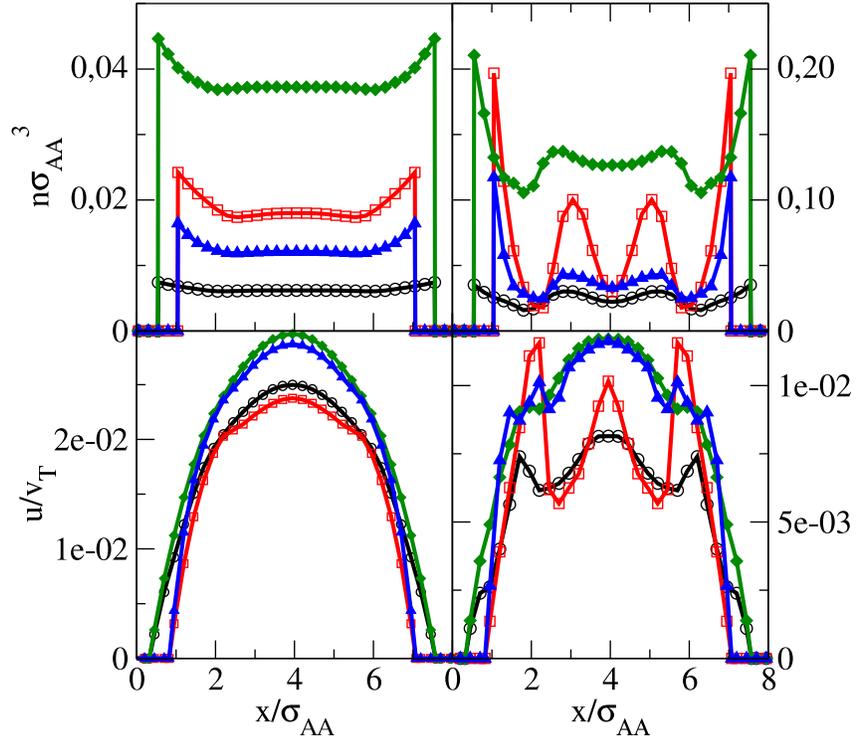}
\caption{Channel of width $H/\sigma_{AA}=7.25.$
The symbols in the four panels correspond to the same conditions as those employed in figure \ref{fig:narrow}.}
\label{fig:wide}
\end{figure}

 In Fig. \ref{fig:wide} we consider a channel with $H=7.25 \sigma_{AA}$. In the low packing region (left panels), there is not significant structure,
 the velocity profiles are Poiseuille-like and display the same trend as in the corresponding panel of Figs. \ref{fig:narrow} and  \ref{fig:medium}. 
 More interesting is the situation at higher packing, shown in the upper right panel. When $\xi_3=0.42$ and  $X_A=0.25$ the large particles
 display four well defined peaks, while the small particles have a flatter structure. Correspondingly, the velocities of the two species
 (lower right panel) are anti-correlated with the density profiles, the largest velocities being attained where the density displays local minima.
Also the velocity profiles at such large packing fraction show an inversion,
that is the larger species has the larger velocity. 
The other case considered  for $\xi_3=0.26$, $X_A=0.75$ displays a less pronounced structure and nearly no oscillations in the 
velocity profiles.

To conclude,  the density profiles are rather sensitive to the composition and
show the characteristic enhancement near  the walls as  packing increases. On the contrary, we have not found a sensitive dependence of
the density profiles on the applied load.
The velocity profiles for moderate packing fractions have shapes reminiscent of
the parabolic Poiseuille-like profiles, with the velocity of the larger species being smaller at low packing , in agreement with the prediction of the toy
model of the previous section.

\subsection{Coarse grained observables.}
\begin{figure}[htb]
\includegraphics[clip=true,width=12.0cm, keepaspectratio,angle=0]{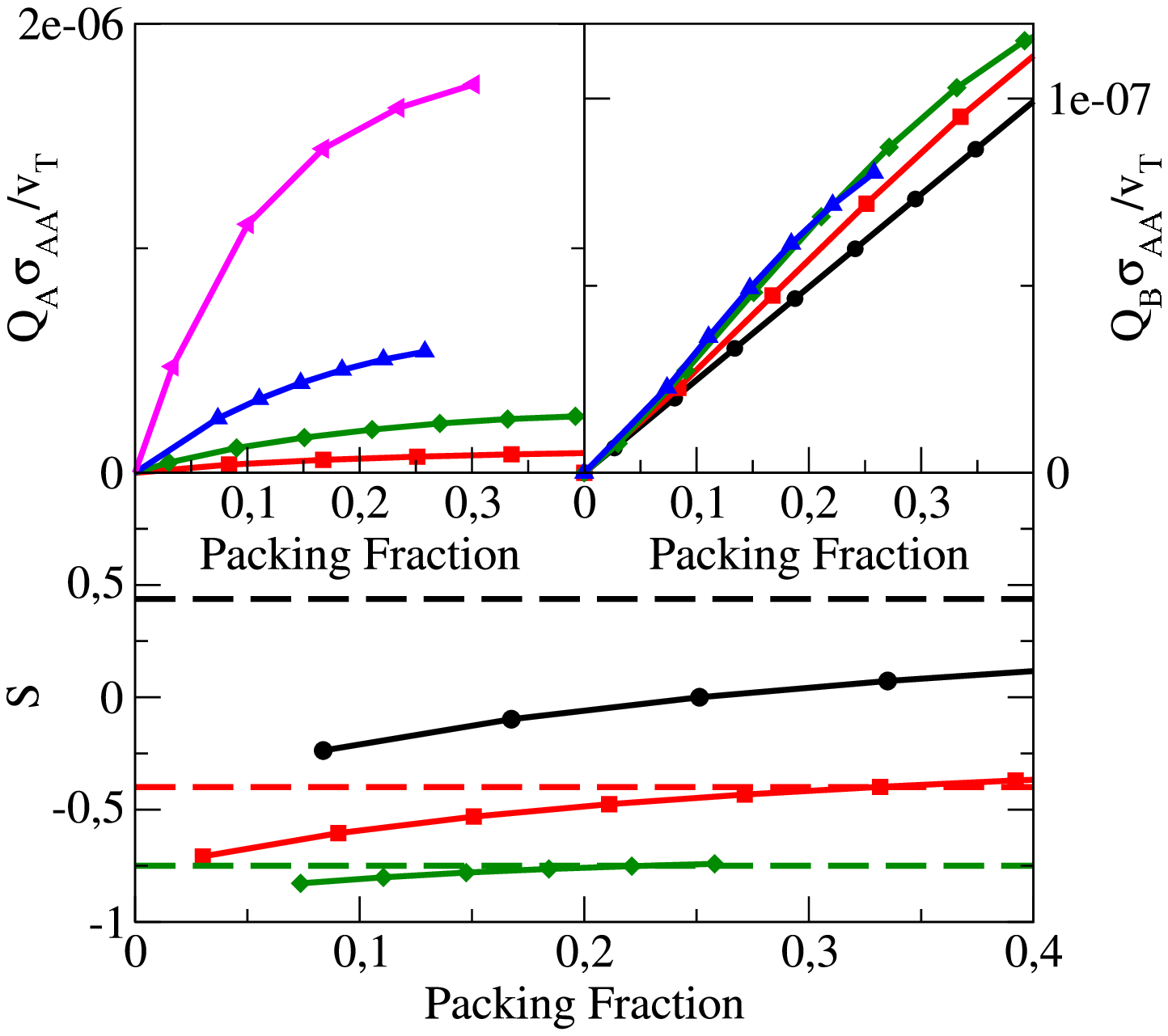}
\caption{Volumetric flow rates for species A (upper left panel) and B (upper right panel) and selectivity function  S (lower panel)  as a function of the packing fraction
for a channel of width $H=2\sigma_{AA}$ and $X_A=0.5$. For the sake of comparison we plotted the lines obtained assuming a simple Poiseuille type of behavior of the
system as explained in the text. }
\label{fig:narrowchannel}
\end{figure}

\begin{figure}[htb]
\includegraphics[clip=true,width=12.0cm, keepaspectratio,angle=0]{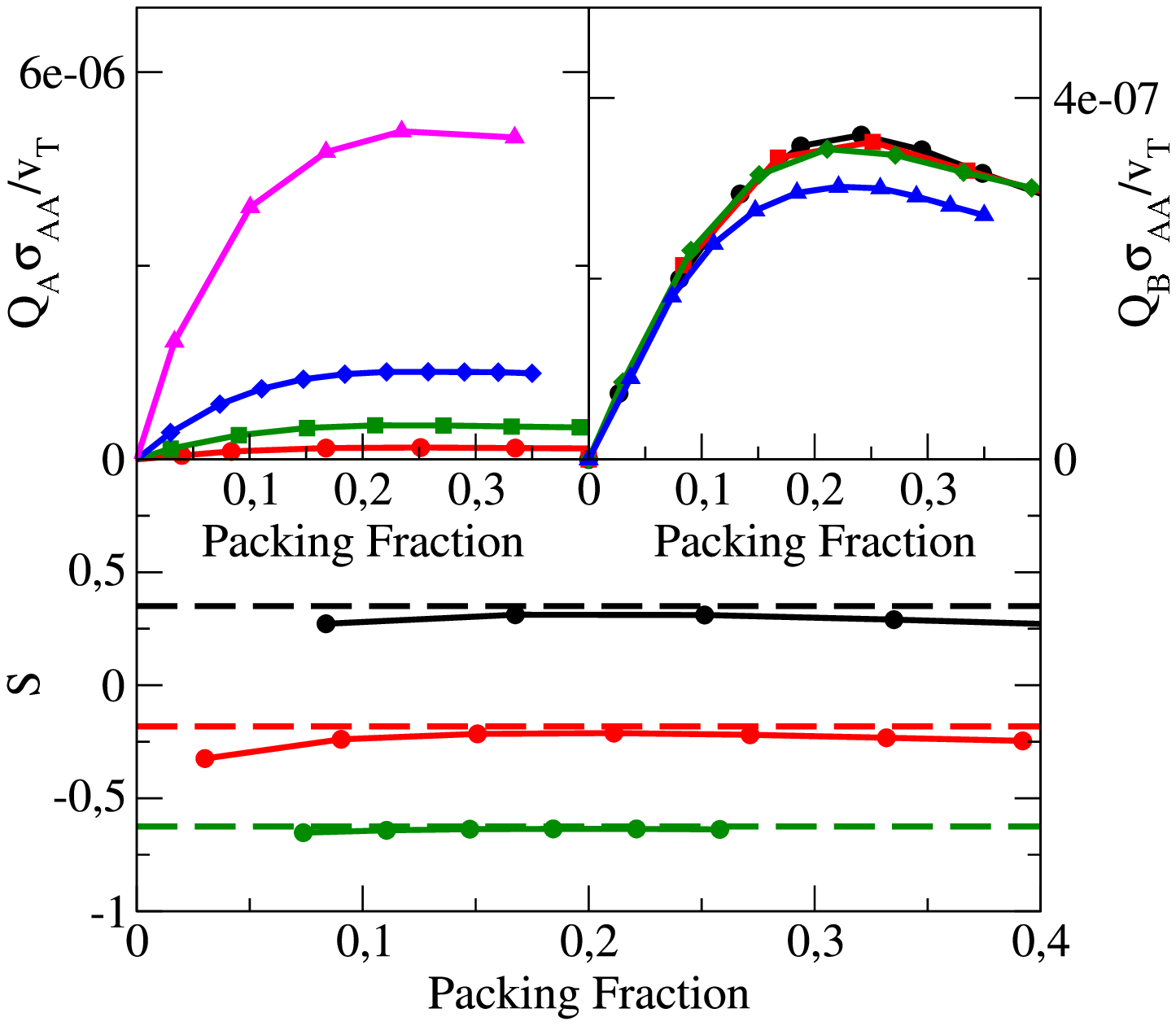}
\caption{Volumetric flow rates for species A (upper left panel) and B (upper right panel) and selectivity function  S (lower panel)  as a function of the packing fraction
for a channel of width $H=2\sigma_{AA}$ and $X_A=0.5$.  For the sake of comparison we plotted the lines obtained assuming a simple Poiseuille type of behavior of the
system as explained in the text. }
\label{fig:mediumchannel}
\end{figure}

Besides considering the microscopic aspect represented by the various profiles, it is of interest to 
analyze some average  properties, such as the volumetric flow of each component and the selectivity, quantities which are 
more easily measurable in experiments.
We define the volumetric flow rate of each species as:
\be
Q^\alpha\equiv {\cal  A} \int_0^H dx   n^\alpha(x)  u^\alpha(x)  
\ee
where ${\cal  A}$ is the area of the plates. In the upper panels of Figs. \ref{fig:narrowchannel} and \ref{fig:mediumchannel}
we display how these quantities vary with the packing fraction, $\xi_3$, for a fixed value of the mixture  composition,
$X_A$. In the lower panel we display how the selectivity, S, defined as
\be
S\equiv \frac{Q^B-Q^A}{Q^A+Q^B}
\ee
varies with the packing at fixed  composition.

We first observe that the volumetric flows at low densities
increase almost linearly with  packing .
In general, the volumetric flow of the small species for equal composition ($X_A=0.5$)  is larger for the small species,
on account of the larger effective volume available.
 If packing increases further, the volumetric flow decreases. Such a feature has the following explanation:
in the low density region,  $Q^\alpha$ increases almost linearly with packing, but as $\xi_3$ becomes larger
it determines a quadratic increase in viscosity and thus a decrease of the velocity of the fluid in the pore.
As a result, there exists an optimal value of the packing for which the volumetric flow is maximum.
The peak shifts towards larger values of the packing as the pore  becomes larger.
Interestingly, such a non-monotonic behavior is captured also by the toy model of section \ref{Channel} .
The results suggest that particles could be separated by their size.
Hydrodynamic chromatography, instead, exploits the Poiseuille velocity profile in channels \cite{rauscher}.

%%%%%%%%%%%%%%%%%%%%%%%%%%%
\section{Conclusions and perspectives}
\label{Conclusions}

To summarize, we have proposed a new theoretical method to study mixtures under confinement
by using concepts derived from both DFT and LBM. The method has been applied to recover bulk properties
such as the shear viscosity, 
and  to study inhomogeneous situations, such as the flow in a slit-like channel of molecular thickness.
Whereas MD simulations provides a complete microscopic picture for 
molecular level mechanisms at the price of a demanding computer time, 
the present method offers an efficient alternative
in terms of CPU-time and high numerical versatility.

We have extended the  kinetic model described in refs.
\cite{Brey,Melchionna2009}   to a multicomponent system.
The formalism enables us to compute not only the transport coefficients
of fluid mixtures, but also to investigate inhomogeneous
dynamical properties.
We have verified that the numerical solution is in agreement with the
values of viscosity predicted by our theory.
In addition, we have considered the  dynamics of a mixture of particles
of different size confined between two parallel plates. 
Under high confinement, different species display different velocity profiles.
We also found non-trivial behavior when the plate separation becomes
so narrow that the harshly repulsive forces among the particles and 
with the confining walls come into play.

We wish to comment that our approach relies on the validity of the local equilibrium 
approximation and is corroborated by the observation that 
 both local thermodynamic  and  hydrodynamic theories appear to be reliable up to 
surprisingly small wavelengths. At smaller scales,
 both conditions break down and a transition to a genuinely molecular regime takes place.
 Compared to previous approaches \cite{LADM,Pozhar,Pozhar2},  our method contains two 
main ideas which in our opinion render it a workable scheme: the simplified kinetic kernel (\ref{breytwo})
and the numerical scheme based on the LBM  techniques.
Very fast numerical solutions can be obtained along these lines and complex geometries can be handled with
great  simplicity and good analytical insight.

In developing the numerical algorithm via a generalized version of the Lattice Boltzmann method, 
we did not take into account temperature variations across the system because we
wanted to maintain the methodology as simple as possible.
The non isothermal extension of the algorithm is doable, but this would require to include
Hermite polynomials up to fourth order (that is, 127 mesh neighbors in the collisional part 
of the algorithm), in order to guarantee numerical stability and physical accuracy.
In the future, we plan  to handle this aspect.

Before concluding we wish to add that several applications of our theory 
can be envisaged to the study of fluids in nanospaces.
Among others we mention just a few:

a)
the spreading of a highly concentrated region in flow conditions and of 
the competition between the  advection and molecular diffusion \cite{Bruus};

b) 
the transport 
of different  molecular species in a channel of varying width
and the influence of the size and/or mass on the dynamical properties \cite{rauscher};

c) 
the entropic, Asakura-Osawa interactions with walls 
and the interplay between advection and diffusion.
When particles of different size are present
one expects that the larger particles experience a larger attraction towards the walls, the so called
effective entropic force \cite{Evans1999,Crocker}.  
At low Reynolds number, the scenario should not be greatly
modified, but the issue is worth to be considered with some care;

d) the dynamics of menisci formed in narrow slits between two coexisting liquid phases
\cite{vanswol};

e) 
driven flows past chemically, physically heterogeneous substrates,
a mechanism which could enhance mixing \cite{Balasz};

f)
since the equilibrium
phase diagram of substances under confinement is
shifted with respect to  bulk conditions, a similar dynamical effect is expected and could be used 
to obtain specific properties \cite{Evans1987};

g) 
cylindrical pores are known  
to lead to strong, non-trivial size selectivity, which is highly 
sensitive to the pore width \cite{Simonejeanpierre}.
It would be very interesting to exploit our model in such a geometry.

%%%%%%%%%%%%%%%%%%%%%%%%%%%%%%%%%%%%%%%%%
\appendix 

\section{The connection with the chemical potential}
In this appendix we give a proof of formula (\ref{potchimico}).
To this purpose let us consider
the r.h.s. of the momentum equation (\ref{momentpartial}) 
for species $\alpha$. 
At equilibrium the l.h.s. of eq. (\ref{momentpartial}) vanishes
together with  the drag and the viscous contributions, so that
we obtain the condition:
\be
k_B T \NN \na(\rr,t)-\na(\rr,t) \FF^{\alpha}(\rr,t)=
\na(\rr,t) \FF^{\alpha,mf}(\rr,t)
\label{born1}
\ee
%to a form which shows the relation of the present approach with Density
%functional theory.

Under the  adiabatic hypothesis, we assume 
the non-equilibrium positional pair correlations to be
given by the equilibrium pair correlations \cite{Tarazona,ArcherEvans}
evaluated when the instantaneous density profiles assume the values
$n^A(\rr,t)$ and $n^B(\rr,t)$. 
The force $ \FF^{\alpha,mf}(\rr)$ can be expressed as
\be
 \FF^{\alpha,mf}(\rr)=-\int d\rr'
 \sum_\beta \gab(\rr,\rr')
\nb(\rr') \NN U^{\alpha\beta}(|\rr-\rr'|) 
\label{born}
\ee
where we have used the
property of the hard sphere potential:
\be
\int d\rr'\nb(\rr') \gab(\rr,\rr') \nabla_r U_{hs}^{\alpha\beta}(|\rr-\rr'|)=-k_B T \int d\rr' \nb(\rr') \gab(\rr,\rr') 
 \delta(|\rr-\rr'|-\sab)
\frac{\rr-\rr'}{|\rr-\rr'|}
\ee
in order to  recast the surface integral 
over the sphere of radius $\sab$ eq.(\ref{potchimico}) into a volume integral.
Substituting the result (\ref{born1}) into eq. (\ref{born}) 
we obtain the  Born-Green-Yvon relation \cite{Evans1}.
We now compare the r.h.s.  of eq. (\ref{born1}) with the exact equilibrium relation 
\bea
&&
k_B T \NN \na(\rr)-\na(\rr) \FF^\alpha(\rr)=
k_B T \na(\rr)\int d\rr'
\sum_\beta c_{\alpha\beta}(\rr,\rr')
\NN' \nb(\rr') 
\nonumber\\
&&
=-\na (\rr) \NN \frac{\delta {\cal F}_{int}}{\delta \na(\rr)}= -\na(\rr)\NN \mu_{int}^\alpha(\rr)
\label{lovett}
\eea
where  ${\cal F}_{int}$ is the non-ideal contribution to the free energy
functional,
$-k_B T c_{\alpha\beta}(\rr,\rr')=
\frac{\delta^2 {\cal F}_{int}}{\delta \na(\rr)\delta \nb(\rr')}$ is the Ornstein-Zernike  
direct correlation function and $-\NN\mu^\alpha_{int}$ is the force arising from the interactions
with the other fluid particles on a particle located at $\rr$ \cite{hansen}.
By equating the r.h.s. terms of  eqs. (\ref{born}) and (\ref{lovett}, we conclude that
eq. (\ref{potchimico}) holds.

Finally, notice that eq. (\ref{lovett}) can be rewritten as
\be
k_B T\NN \ln \na(\rr)+\NN \mu^\alpha_{int}(\rr)=\FF^\alpha(\rr)
\ee
that is, in thermal and mechanical equilibrium the net force on a fluid particle
at position $\rr$ must vanish.
This occurs  when the entropic force, represented by the first term, being
the resultant of the
attractive and repulsive forces exerted on a molecule at $\rr$ from all the other molecules
(the second term on the l.h.s.) and the external force (on the r.h.s.) compensate exactly.

%%%%%%%%%%%%%%%%%%%%%%%%

%%%%%%%%%%%%%%%%%%%%%%%

\section{The discretization procedure}

A crucial goal of the present work is to achieve the solution of the  coupled integro-differential 
equations for non-ideal 
mixtures by means of a suitable numerical scheme.
In this work, we consider the Lattice Boltzmann method as the reference framework that enables us
to solve the kinetic equation in bulk conditions and under confinement, in particular when the confining 
geometry is highly non-trivial.
The main asset of the LBM is to work directly with the distribution functions by accomplishing 
a spectral decomposition of the distribution in velocity space via a Hermite representation. 
In this way, the three-dimensional velocity space is reduced to a handful of sampling points
and the three-dimensional real space is discretized over a Cartesian mesh. The resulting distribution 
function is transformed into a reduced set of populations.
The phase space evolution is thus rewritten as an efficient updating algorithm where
the streaming operator has the form of a forward Euler updating step.
The LBM approach has been previously employed to solve the collisional dynamics of the single-component 
hard-sphere fluid and provided robust solutions up to packing fraction $<0.35$  \cite{Melchionna2009}.

The LBM is a very general framework to handle the evolution of generic kinetic equations and
does not depend in any specific way on the form of the collisional kernel \cite{LBgeneral}.
We exploit this generality to solve the collisional dynamics encoded by eq. (\ref{breytwo}),
where non-local structural forces appear as surface integrals, and retain the simple LBM picture
of updating the populations. In addition, the BGK kernel of eq.  (\ref{breytwo}) also depends on the Hermite 
representation via the equilibrium distribution. A second-order expansion of the distribution function, 
corresponding to a Hermite representation to second order, is usually employed in one-component 
LBM implementations \cite{shanchen,martys,heluo,shanhe}.

The distribution function of each species is discretized in velocity space and the continuous phase space 
distribution is replaced by a $Q$-dimensional array, $\fa(\rr,\vv,t)\rightarrow \fa_p(\rr,t)$
where $p=1,\dots,Q$ labels discrete velocities ${\bf c}_p$.
The $Q$ discrete velocities connect neighboring mesh points  on a lattice and mirror
the hop of particles between mesh points, generally augmented by a null 
vector ${\bf c}_{0}$ accounting for particles at rest.
The form of the lattice velocities ${\bf c}_p$ depends on the order of accuracy of the method
and reflects into the required Hermite order \cite{shanyuanchen}.

In essence, the LBM dynamics is achieved by a rearrangement of populations over spatial shifts via an explicit update 
arising from the following approximation to the streaming operator,
\be
\partial_{t}f_{p}(\rr,t)+\cc_{p}\cdot\partial_{\rr}f_{p}(\rr,t) \simeq
\frac{f_{p}(\rr+\cc_{p}\Delta t,t+\Delta t)-f_{p}(\rr,t)}{\Delta t}
\ee
where $\Delta t$ is the time-step. We choose the mesh classified as D3Q19, that is, containing $19$ discrete 
velocities connecting first and second neighbors of a cubic lattice and overall achieving second-order 
Hermite accuracy. 
We choose equal masses for the hard-sphere species so that the thermal velocity is the same for both species
and given by $v_{T}=1/\sqrt{3}$. An extension of the present method to different masses can be easily obtained
by using the method described in ref. \cite{Zhaoli}.
The updating step for the populations is compactly written as
\be
f^\alpha_{p}(\rr+{\bf c}_{p},t+\Delta t) = \bar{f}^\alpha_{p}(\rr,t)
\ee
where  ${\bar f}^\alpha_{p}(\rr,t)$ is the post-collisional population of species $\alpha$, 
containing the effect of the BGK relaxation and the hard-core collisions, written as  
\be
{\bar f}^\alpha_{p}(\rr,t)= {\bar f}^{bgk,\alpha}_{p}(\rr,t) + {\bar f}^{hs,\alpha}_{p}(\rr,t)
\ee

The BGK contribution to the post-collisional population has the form
\be
{\bar f}^{bgk,\alpha}_{p}(\rr,t) = 
(1 - \omega \Delta t)f^\alpha_{p}(\rr,t) 
+ \omega \Delta t\psi^\alpha_{\perp,p}(\rr,t)
\ee
where 
\be
\psi^\alpha_{\perp,p}({\bf r },t) =
w_{p}n^\alpha({\bf r},t)
\left[
1 + 
\frac{c_{pi}u^\alpha_{i}({\bf r },t)}{v_{T}^{2}} + 
\frac{(c_{pi}c_{pj}-v_{T}^{2}\delta_{ij})
\left(
v_T^2 \delta_{ij}
+ 2 u^\alpha_i({\bf r},t)u_{j}({\bf r},t)
-u_i({\bf r},t)u_{j}({\bf r},t)\right)}{2v_{T}^{4}}
\right]
\ee
and
\be
\psi^\alpha_{p}(\rr,t) =
w_{p}n^\alpha(\rr,t)
\left[
1 + \frac{c_{pi}u_{i}(\rr,t)}{v_{T}^{2}} + 
\frac{(c_{pi}c_{pj}-v_{T}^{2}\delta_{ij})u_{i}(\rr,t)u_{j}(\rr,t)}{2v_{T}^{4}}
\right].
\label{b6}
\ee
The $w_p$ are the $Q$ weights normalized to unity that arise from the Hermite expansion \cite{LBgeneral}.
Formula (\ref{b6})
is a low-Mach ($O[Ma^{3}]$) expansion of the local Maxwellian corresponding to the 
barycentric velocity for a mixture of equal masses
\be
{\bf u}(\rr,t) = { \sum_\alpha n^\alpha {\bf u}^\alpha(\rr,t) \over \sum_\alpha n^\alpha(\rr,t) }
\ee
with
\bea
n^\alpha(\rr,t) &=& \sum_p f_p^\alpha(\rr,t) \nonumber \\
n^\alpha(\rr,t) {\bf u}^\alpha(\rr,t) &=& \sum_p \cc_p f_p^\alpha(\rr,t)
\eea

The hard-sphere post-collisional contribution is 
\be
{\bar f}^{hs,\alpha}_{p}(\rr,t) = -w_{p} 
\left[
\frac{c_{pi}C^{(\alpha)}_{i}(\rr,t)}{v_{T}^{2}}+\frac{(c_{pi}c_{pj}-v_{T}^{2}\delta_{ij})u_{i}(\rr,t)C^{(\alpha)}_{j}(\rr,t)}{v_{T}^{4}}
\right].
\ee
The collisional integral is approximated via the following real-space quadrature 
\bea
{\bf C}^{(\alpha)}(\rr,t) = 
\sum_\beta
{\bf C}^{\alpha\beta}(\rr,t) 
=
&&
- 4\pi \sum_\beta
\sab^2 
\sum_q
W_q \bk_q
g_{\alpha\beta}(\rr+ {1\over 2}\sab \bk_q,t)
n_{\alpha}(\rr,t) 
n_{\beta}(\rr+\sab\bk_q,t) \times \nonumber\\
&&
\Bigl( v_T^2 - 2 \sqrt{\frac{2 v_T^2}{\pi} } \bk_q\cdot
[\uub(\rr+\sab\bk_q,t)-\uua(\rr,t)]
\Bigl)
\label{cintegral}
\eea
where $\bk_q$ are the nodes of the quadrature over a spherical surface \cite{abramovitzstegun}
 and $W_q$ the
associated quadrature weights. We choose a 18-points quadrature that is exact for quadratic integrands. 
The quadrature nodes on the unit sphere are the six on-axis points 
$(\pm 1,0,0)$, $(0,\pm 1,0)$ and $(0,0,\pm 1)$
and twelve points along the diagonals $(\pm \sqrt{2},\pm \sqrt{2},0)$, 
$(\pm \sqrt{2},0, \pm \sqrt{2})$ and $(0,\pm \sqrt{2},\pm \sqrt{2})$.
Correspondingly, the quadrature weights are $W_q=1/15$ for the on-axis points and 
$W_q=1/30$ for the diagonal points.
It should be borne in mind that the quadrature points do not necessarily fall 
on the mesh. For instance, if the the HS radius is an integer and one considers
intraspecies collisions, there are only six quadrature nodes that fall on the mesh 
while twelve nodes along the diagonals do not.
In order to sample the fluid density and velocity on off-mesh points, we employ a trilinear 
interpolation scheme \cite{NUMRECIPES}.
It is clear that the accuracy of the collisional integrals improve for larger HS diameters, 
since the eight mesh points employed in the trilinear interpolation fall closer to the 
spherical surface. The same type of consideration applies for the interspecies integral and, 
in all circumstances, for the midpoint rule employed
to estimate the radial distribution function via the Carnahan-Starling expression eq. (\ref{carnahan}).
In our numerical applications,
we applied a trilinear interpolation 
for any off-mesh quadrature nodes and midpoints for the radial distribution function, 
and chose HS diameters to be even numbers in order to  minimize the number of off-mesh interpolations.

In solving eq. (\ref{breytwo}) with the LBM approach, we need to control the numerical error introduced by the
discretization procedure over velocity, space and time variables. It is well-known that 
for one-component, uncorrelated fluids LBM bears three types of error as compared to the 
analytical solution of the incompressible 
Navier-Stokes equation. The first error depends on the mesh spacing $\Delta x$ and decays 
quadratically with $\Delta x$. In presence of hard boundaries, and in particular with curved 
fluid-wall interfaces that are approximated as staircase geometries, the accuracy degrades 
to linear dependence with the mesh spacing.
The second type of error depends quadratically on the time-step $\Delta t$. Finally, the so-called 
compressibility error arises from the fact that LBM does not  exactly enforce a non-zero divergence 
of the velocity field, with a resulting error that depends quadratically on the Mach number. 
As a side effect, the time-step error also degrades linearly with $\Delta t$. These three types 
of error can be made arbitrarily small by changing $\Delta x$, $\Delta t$ and the mass unit 
while keeping fixed the physical value of the mesh spacing, kinematic viscosity and mass density.
In order to minimize the composite error, the mesh spacing is typically the parameter that is reduced, 
by negotiating with the numerical effort needed to resolve local morphological details and flow pattern.
A concomitant effect of the space-time discretization is the effective reduction of the kinematic 
viscosity arising from a negative viscosity of numerical origin, $-\frac{v_T^2\Delta t}{2}$, that
depends on the specific form of the lattice employed, such as D3Q19 in our case.
The numerical contribution to viscosity is evaluated via a Chapman-Enskog analysis
with the outcome that the effective viscosity can be controlled to arbitrary accuracy \cite{LBgeneral}.
For the case of correlated dynamics, the viscosity due to the finite-time propagation enters 
with the same numerical value of the uncorrelated case, as shown in our previous 
publication \cite{Melchionna2009}.

The extension of the LBM to non-ideal fluids, such as the binary mixture described in this paper,
carries the same types of error as for the uncorrelated case. The BGK contribution to the mixture 
dynamics, encoded by the kernel eq. (\ref{breytwo}), generates the same type of numerical error as for the 
single-component case.
The issue of resolving the atomic-scale structural correlations critically depends on the rapidly 
varying oscillations of the density profiles, as much as on the current and higher moments profiles. 
The profiles need to be resolved at spatial scale smaller than the molecular radius 
$\Delta x < \min_\alpha \sigma_{\alpha\alpha}/2$. When confronting with experiments performed on
nanoscopic systems, the computational load needed to solve the kinetic equation requires a number 
of mesh points that scales as $\sim \sigma_{\alpha\alpha}^3$.
The same argument applies when resolving the temporal evolution, with a time-step
that must be smaller than the typical collision time, $\Delta t \le \min_\alpha \sigma_{\alpha\alpha} / v_T$. 
Regarding the additional compressibility arising in LBM, the error plays a minor 
role as compared to the highly compressible nature of fluids at molecular scale. 
The low-Mach numbers often employed in nanofluidics, condensed-matter conditions further alleviates 
this problem.

In addition to the standard error terms, the computation of the collisional integrals eq. (\ref{cintegral}) 
introduces a further source of error. 
It is a simple exercise to show that for a uniform system the diagonal points in the quadratures 
produce the collisional contribution to viscosity. Therefore, the numerical error on 
the collisional viscosity are mostly due to the off-lattice location of the quadrature points
and the employed trilinear interpolation scheme.
The level of accuracy of viscosity
can be controlled in a systematic way and the theoretical curve can be recovered to an excellent
level. Our numerical results indicate that, once the molecular spatial scale is
resolved to account for the microscopic flow patterns, the error due to the quadratures 
becomes negligible.

Lattice nodes adjacent to the walls obey collision rules different from
those characterizing bulk nodes. We have adopted the so called
bounce-back collision rule which states that the velocity
 of a particle incident on a wall is reversed after the collision.

\begin{acknowledgements}
U.M.B.M. acknowledges the support received from the European Science Foundation (ESF) for the activity entitled 'Molecular Simulations in Biosystems and Material Science.
\end{acknowledgements}
%%%%%%%%%%%%%%%%%%%%%%%%%
%%%%%%%%%%%%%%%%%%%%%%%%%%%

%\end{document}
%%%%%%%%%%%%%%%%%%% FIGURES
%%% ------------------------------------- FIG 1
%----------------------------- Bulk shear viscosity--------------------------------------%%%%%%%%%%%%%%%%%%%%%%%%%%
%----------------------------- Fig.2 --------------------------------------
%                            CARTOON
%%%%%%%%%%%%%%%%%%%%%%%\end{document}
%----------------------------- Fig.3 --------------------------------------
%%%%%%%%                   TOY MODEL PROFILES

%%%% --------------- FIG 4
%----------------------------- Fig. 4 Narrow channel --------------------------------------

%----------------------------- Fig. 5 Medium channel --------------------------------------

%----------------------------- Fig. 6 Wide channel --------------------------------------

%----------------------------- Fig. 7 Narrow channel VOLUMETRIC--------------------------------------
%----------------------------- Fig. 8 Wide  channel VOLUMETRIC --------------------------------------

\end{document}